\documentclass[aps,prd,preprint,preprintnumbers,unsortedaddress,superscriptaddress,showpacs,nofootinbib]{revtex4-1}

\pdfoutput=1
\usepackage{bm}
\usepackage{graphicx}
\usepackage{amsmath}
\usepackage{amsfonts}
\usepackage{amssymb}
\usepackage{color}
\usepackage[section]{placeins}
\usepackage{booktabs}
\usepackage[colorlinks, citecolor=blue,anchorcolor=red,menucolor=red, linkcolor=red,filecolor=red,runcolor=red,urlcolor=blue,frenchlinks=red]{hyperref}

\begin{document}
\arraycolsep1.5pt

\title{ $\tau^- \to \nu_{\tau} M_1 M_2$,  with $M_1, M_2$ pseudoscalar or vector mesons}


\author{L.~R.~Dai}
\email{dailr@lnnu.edu.cn}
\affiliation{Department of Physics, Liaoning Normal University, Dalian 116029, China}
\affiliation{Departamento de F\'isica Te\'orica and IFIC, Centro Mixto Universidad de Valencia-CSIC,
Institutos de Investigac\'ion de Paterna, Aptdo. 22085, 46071 Valencia, Spain
}

\author{R. Pavao}
\email{rpavao@ific.uv.es}
\affiliation{Departamento de F\'isica Te\'orica and IFIC, Centro Mixto Universidad de Valencia-CSIC,
Institutos de Investigac\'ion de Paterna, Aptdo. 22085, 46071 Valencia, Spain
}

\author{S. Sakai}
\email{shuntaro.sakai@ific.uv.es}
\affiliation{Departamento de F\'isica Te\'orica and IFIC, Centro Mixto Universidad de Valencia-CSIC,
Institutos de Investigac\'ion de Paterna, Aptdo. 22085, 46071 Valencia, Spain
}

\author{E.~Oset}
\email{oset@ific.uv.es}
\affiliation{Departamento de F\'isica Te\'orica and IFIC, Centro Mixto Universidad de Valencia-CSIC,
Institutos de Investigac\'ion de Paterna, Aptdo. 22085, 46071 Valencia, Spain
}

\date{\today}
\begin{abstract}
 We perform a calculation of the $\tau^- \to \nu_{\tau} M_1 M_2$,  with $M_1, M_2$ either pseudoscalar or vector mesons using  the basic weak interaction
 and  angular momentum algebra to  relate the different processes. The formalism also leads to a different interpretation of the role played by $G$-parity in these decays. We also observe that,
 while $p$-wave $M_1 M_2$ production is compatible with chiral perturbation theory and experiment, $VP$ and $VV$ $p$-wave production is clearly incompatible
 with experiment and we develop the  formalism  also in this case. We compare our results with experiment and make predictions for unmeasured  decays,
 and we show the value of these reactions, particularly if the $M_1 M_2$ mass distribution is measured, as a tool to learn about the meson-meson interaction
 and the nature of some resonances, coupling to two mesons, which are produced in such decays.

\end{abstract}
\maketitle

\section{Introduction}
\label{intro}
Tau decays have been instrumental to learn about weak interaction as well as strong interaction affecting the hadrons produced on $\tau^-$  hadronic decay \cite{isgur89,npb84,zpc90,PRe88,Rep2006,PRe2010}.
$\tau^-$ decays into $\nu_{\tau}$ and a pair of mesons make up for a sizeable fraction of the $\tau^-$ decay width \cite{pdg}.  Several modes are well measured, as
 $\tau^- \to \nu_{\tau} K^{0} K^{-}$ \cite{Ryu} \footnote{We mention explicitly the most recent experiments. The full information can be obtained in Ref. \cite{pdg}.},
$\tau^- \to \nu_{\tau} \pi^{-} \bar{K}^{0}$ \cite{Epifanov}, $\tau^- \to \nu_{\tau} \pi^{-} \omega $ \cite{Buskulic1,Buskulic2},
$\tau^- \to \nu_{\tau} K^{*0} K^{-}$ \cite{Barate},   $\tau^- \to \nu_{\tau} \eta K^{*-} $ \cite {Inami},  $\tau^- \to \nu_{\tau} K^{-} \omega $ \cite{Arms},
$\tau^- \to \nu_{\tau} \pi^{0} \rho^{-}$ \cite{Barate2}, $\tau^- \to \nu_{\tau} \pi^{-} K^{*0} $ \cite{Barate},
$\tau^- \to \nu_{\tau} \pi^{-} \phi$  \cite{Aubert}, $\tau^- \to \nu_{\tau} K^{-} \phi$ \cite{Aubert},    $\tau^- \to \nu_{\tau} \eta K^{-}$ \cite{delAmo}.
As we can see, there are modes with two pseudoscalar mesons and also modes with pseudoscalar-vector. Surprisingly,  there are no vector-vector modes reported in the PDG \cite{pdg}.
Certainly the large mass of the vector mesons leaves small phase space for the decay, but modes like $\rho^{0} \rho^{-}$, $\rho^{-} \omega$, $K^{*-} \rho^{0}$, $K^{*-} \omega$,
$\bar{K}^{*0} \rho^{-} $ are kinematically possible, and  even  $K^{*0}K^{*-}$  considering the width of $K^{*}$. One may wonder whether there is some fundamental reason for
this experimental fact. Actually, in as much as  the pseudoscalar and vector mesons differ only by the spin arrangement of the quarks,  it should be possible to relate the
rates of decay for two pseudoscalar mesons and the related pseudoscalar-vector or  vector  modes, for instance,
$\tau^- \to \nu_{\tau} K^{0} K^{-}, \nu_{\tau} K^{0} K^{*-},  \nu_{\tau}  K^{*0} K^{-}, \nu_{\tau}  K^{*0} K^{*-}$.
Based on the basic dynamics of the weak interaction and using the $^3P_0$ model \cite{micu,oliver,close} to hadronize into two mesons the
primary $q\bar{q}$ state formed, we relate the widths of such decay modes.

One interesting point concerning $\tau^-$ mesonic decays is the issue of charge symmetry discussed in Ref. \cite{Weinberg} and the classification of the weak interaction into
first and second class currents.  The issue, with suggestions of experiments, is retaken  in Refs. \cite{Leroy,Escribano,roig}.
One of the interesting reactions is the  $\tau^- \to \nu_{\tau} \pi^{-} \eta (\eta^{\prime})$, which according to that classification is forbidden by $G$-parity,
and efforts are made to go beyond the standard model to get contributions to this decay modes  \cite{Escribano,roig}.

The $G$-parity plays indeed an important role in these reactions and in this paper we offer a new perspetive into this issue. We shall see that  $G$-parity
for the non strange mesons plays an important role and the rules are different for pseudoscalar-pseudoscalar ($PP$) pseudoscalar-vector ($PV$) or vector-vector ($VV$)
production.  But an extension of these rules appears also in the strange sector for the  $\tau^- \to \nu_{\tau} K^- \eta (\eta^{\prime}),\nu_{\tau} K^{*-} \eta (\eta^{\prime})$
reactions.

    We make a thorough study of all possible Cabibbo-favored and Cabibbo-suppressed reactions and compare with present available data.

\section{Formalism}
\label{sec:form}
\begin{figure}
	\begin{center}
		\includegraphics[width=0.48\textwidth]{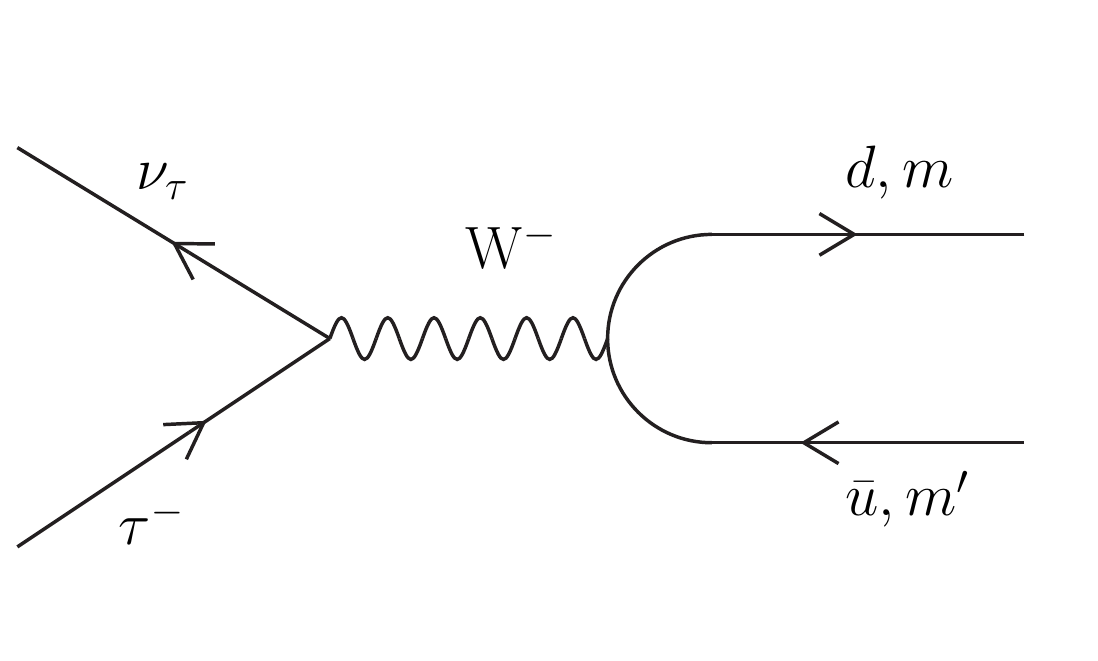}	
	\end{center}
	\caption{\label{fig:tau} Elementary $\tau^- \to \nu_{\tau} d \bar{u}$ diagram. The labels $m,m^{\prime}$ stand for the third component of spin of the quarks. }
\end{figure}

\begin{figure}
	\begin{center}
\includegraphics[width=0.55\textwidth]{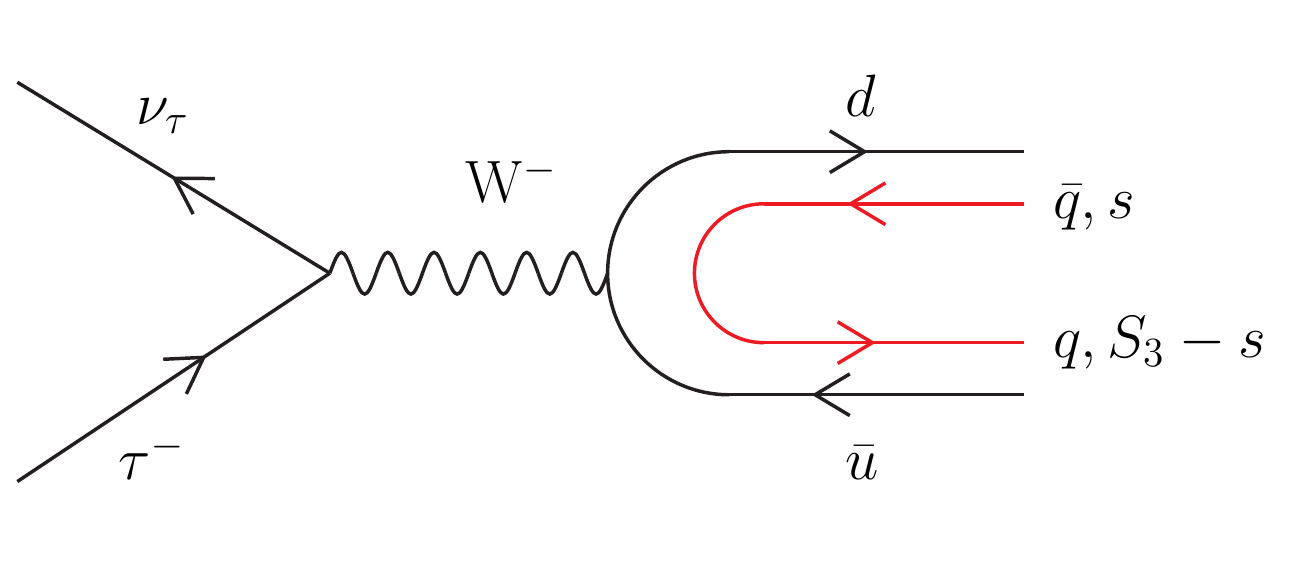}	
	\end{center}
\caption{\label{fig:tauh} Hadronization  of the primary $d\bar{u}$ pair to produce two mesons, $s$ is the third component of the spin of $\bar{q}$ propagating as a particle, while
$S_3-s$ is the third component of the spin of $q$ , where  $S_3$ is the third component of the total  spin $S$ of $\bar{q}q$. }
\end{figure}

The first step is to look at the $\tau^- \to \nu_{\tau} q \bar{q}$  decay depicted in Fig. \ref{fig:tau} for the Cabibbo-favored $d\bar{u}$  production.
We obtain the Cabibbo-suppressed mode substituting the $d$ quark by an $s$  quark. However, we are interested in the production of two mesons, not just one,
as it would come from the mechanism of  Fig. \ref{fig:tau}  when $q \bar{q}$  merge into a meson. The procedure to produce two mesons is hadronization
by creating a new $q \bar{q}$  pair with the quantum numbers of the vacuum. This is depicted in  Fig. \ref{fig:tauh}.

  It is easy and relevant to see how two mesons appear, and in which order, to see the relevance of the $G$-parity in the reactions. For this purpose,
and looking only at the flavor components, we proceed as follows \cite{zou,liang}: we introduce the matrix $M$
\begin{eqnarray}
M = \left(
           \begin{array}{ccc}
             u\bar u & u \bar d & u\bar s \\
             d\bar u & d\bar d & d\bar s\\
             s\bar u & s\bar d & s\bar s\\
           \end{array}
         \right)\nonumber\,,
\end{eqnarray}
and when we  do the hadronization of $d\bar{u}$ we get
\begin{eqnarray}
d\bar{u} \to \sum\limits_{i=1}^{3} d~ \bar{q}_i~ q_i ~\bar{u} =M_{2i} ~M_{i1}=(M \cdot M)_{21} \nonumber\, .
\end{eqnarray}
And now we write the $M$ matrix in terms of pseudoscalar or vector mesons
\begin{eqnarray}\label{eq:P}
P=\begin{pmatrix}
\frac{\pi^0 }{\sqrt{2}} +\frac{\eta}{\sqrt{3}}  +\frac{\eta'}{\sqrt{6}}   & \pi^+ &  K^+ \nonumber\,\\
\pi^- & -\frac{\pi^0 }{\sqrt{2}} +\frac{\eta}{\sqrt{3}}  +\frac{\eta'}{\sqrt{6}}  & K^0 \nonumber\,\\
K^- & \bar{K}^0 & - \frac{\eta}{\sqrt{3}}+\frac{2\eta'}{\sqrt{6}}
\end{pmatrix},
\end{eqnarray}
where the standard mixing of $\eta$ and $\eta'$ has been assumed~\cite{bramon},
\begin{equation}
\label{eq:V}
 V_{\mu} =
\left(
\begin{array}{ccc}
\frac{\rho^0}{\sqrt{2}}  + \frac{\omega }{\sqrt{2}} & \rho^+ & K^{* +} \\
\rho^- & -\frac{\rho^0}{\sqrt{2}}  + \frac{\omega}{\sqrt{2}}  & K^{* 0}  \\
K^{* -} & \bar{K}^{* 0} & \phi  \\
\end{array}
\right)_{\mu} \nonumber\,.
\end{equation}
Then $M^2$ becomes $PP, PV,VP,VV$ and it is important to keep the order of the mesons. Thus we get
\begin{eqnarray}\label{eq:PP21}
(P \cdot P)_{21}&=& \pi^-\left(\frac{\pi^0}{\sqrt{2}}+ \frac{\eta}{\sqrt{3}} + \frac{\eta^{\prime}}{\sqrt{6}} \right)+\left(-\frac{\pi^0}{\sqrt{2}}+ \frac{\eta}{\sqrt{3}} + \frac{\eta^{\prime}}{\sqrt{6}} \right) \pi^- + K^0 K^- \nonumber\,\\
&=&\left( \pi^- \frac{\pi^0}{\sqrt{2}}-\frac{\pi^0}{\sqrt{2}} \pi^- \right)+ \pi^- \left(\frac{\eta}{\sqrt{3}}+ \frac{\eta^{\prime}}{\sqrt{6}} \right)
+\left(\frac{\eta}{\sqrt{3}}+ \frac{\eta^{\prime}}{\sqrt{6}} \right)\pi^{-} + K^0 K^- \,.
\end{eqnarray}

We shall see later that it is precisely the combination of $\pi^- \eta (\eta^{\prime})$ and $ \eta (\eta^{\prime}) \pi^-$ that appears in Eq. \eqref{eq:PP21} what makes the
$\tau^- \to \nu_{\tau} \pi^{-} \eta $ decay $G$-parity  forbidden, while the $\pi^- \pi^0$, $\pi^0 \pi^- $ combination gets reinforced by the relative sign in
Eq. \eqref{eq:PP21}.

Similarly, we obtain
\begin{eqnarray}\label{eq:PV21}
(P \cdot V)_{21}=\pi^- \left( \frac{\rho^0}{\sqrt{2}}+\frac{\omega}{\sqrt{2}}  \right)+
\left(-\frac{\pi^0}{\sqrt{2}}+\frac{\eta}{\sqrt{3}}+ \frac{\eta^{\prime}}{\sqrt{6}} \right)\rho^{-} + K^0 K^{*-} \, ,
\end{eqnarray}
\begin{eqnarray}\label{eq:VP21}
(V \cdot P)_{21}=\rho^- \left(\frac{\pi^0}{\sqrt{2}}+\frac{\eta}{\sqrt{3}}+ \frac{\eta^{\prime}}{\sqrt{6}} \right)
 + \left( -\frac{\rho^0}{\sqrt{2}}+\frac{\omega}{\sqrt{2}}  \right)\pi^- + K^{*0}K^- \, .
\end{eqnarray}

We see again that $\pi^{-} \rho^0 $ appears as $\pi^{-} \rho^0 $  or $ -\rho^0 \pi^{-} $, $\pi^{-}\omega$  and $\omega \pi^{-}$ and
$(\frac{\eta}{\sqrt{3}}+ \frac{\eta^{\prime}}{\sqrt{6}})\rho^-$  with $\rho^- (\frac{\eta}{\sqrt{3}}+ \frac{\eta^{\prime}}{\sqrt{6}})$.
Once again, we shall see that the order matters in the $G$-parity conservation.

Thirdly, for the $VV$ combination we get
\begin{eqnarray}\label{eq:VV21}
(V \cdot V)_{21}=\left(\rho^- \frac{\rho^0}{\sqrt{2}}-\frac{\rho^0}{\sqrt{2}}\rho^- \right)+\left(\rho^- \frac{\omega}{\sqrt{2}}+\frac{\omega}{\sqrt{2}}\rho^- \right)
 + K^{*0}K^{*-} \, ,
\end{eqnarray}
with, again, relevant signs between the $\rho^- \rho^0$, $\rho^0 \rho^-$ and $\rho^- \omega, \omega \rho^-$ components.

Replacing the $d$ quark by an $s$ quark we get the Cabibbo-suppressed modes. The hadronization  leads to
\begin{eqnarray}
s\bar{u} \to \sum\limits_{i=1}^{3} s~ \bar{q}_i~ q_i ~\bar{u} =M_{3i} ~M_{i1}=(M \cdot M)_{31} \nonumber\, ,
\end{eqnarray}
with the results
\begin{eqnarray}\label{eq:PP31}
(P \cdot P)_{31}=K^- \frac{\pi^0}{\sqrt{2}}+ \bar{K}^0 \pi^- + \left( K^-\frac{\eta}{\sqrt{3}}-\frac{\eta}{\sqrt{3}}K^- \right)
+ \left(K^- \frac{\eta^{\prime}}{\sqrt{6}}+\frac{2\eta^{\prime}}{\sqrt{6}} K^- \right)\, ,
\end{eqnarray}
\begin{eqnarray}\label{eq:PV31}
(P \cdot V)_{31}=K^- \left(\frac{\rho^0}{\sqrt{2}}+\frac{\omega}{\sqrt{2}}\right)+\bar{K}^0 \rho^-
+\left(-\frac{\eta}{\sqrt{3}}+\frac{2\eta^{\prime}}{\sqrt{6}} \right)K^{*-} \, ,
\end{eqnarray}
\begin{eqnarray}\label{eq:VP31}
(V \cdot P)_{31}=K^{*-} \left(\frac{\pi^0}{\sqrt{2}}+\frac{\eta}{\sqrt{3}}+\frac{\eta^{\prime}}{\sqrt{6}}\right)+\bar{K}^{*0}\pi^- +\phi K^{-}  \, ,
\end{eqnarray}
\begin{eqnarray}\label{eq:VV31}
(V \cdot V)_{31}=K^{*-} \left(\frac{\rho^0}{\sqrt{2}}+\frac{\omega}{\sqrt{2}}\right)+\bar{K}^{*0}\rho^- +\phi K^{*-}  \, .
\end{eqnarray}

Interestingly, even if here we do  not have  $G$-parity states, we have also some states appearing in different order, as $K^{-}\eta$, $\eta K^{-}$
and $K^{-}\eta^{\prime}$, $\eta^{\prime} K^{-}$ in $PP$ and $\eta K^{*-}$, $ K^{*-} \eta$, $\eta^{\prime} K^{*-}$, $ K^{*-} \eta^{\prime}$ in $PV$, $VP$.
This has also consequences, similar to those leading to  $G$-parity selection rules, as we shall see.

\subsection{Weak interaction}
\label{sec:weak}

We shall not worry about the global normalization and concentrate only on the relationship of the different decay modes discussed before. Then the weak
interaction is given by
\begin{eqnarray}
H= \mathcal{C} L^\mu Q_\mu,
\end{eqnarray}
with $\mathcal{C}$ containing weak interaction constants and radial matrix elements that we shall see later on, where $L^\mu$ is the leptonic current
\begin{eqnarray}
L^\mu=\langle \bar u_\nu |\gamma^\mu- \gamma^\mu\gamma_5| u_\tau\rangle,
\end{eqnarray}
and $Q_\mu$ the quark current
\begin{eqnarray}
Q^\mu=\langle \bar u_d|\gamma^\mu-\gamma^\mu\gamma_5|v_{\bar u}\rangle.
\end{eqnarray}
As is usual in the evaluation of decay widths to three final particles, we evaluate the matrix elements in the frame where the two mesons system is at rest.
For the evaluation of the matrix element $Q_\mu$ we assume that the quark spinors are at rest in that frame and we have
in the Itzykson-Zuber normalization \cite{itzy}
\begin{equation}\label{eq:wfn}
u_r=
\left(
\begin{array}{c}
\chi_r\\ 0
\end{array}
\right), \qquad
v_r=
\left(
\begin{array}{c}
0\\ \chi_r
\end{array}
\right),\qquad
\chi_1=
\left(
\begin{array}{c}
1\\ 0
\end{array}
\right),\qquad
\chi_2=
\left(
\begin{array}{c}
0\\ 1
\end{array}
\right),
\end{equation}
with  the $\gamma^\mu$ matrices,
\begin{equation}\label{eq:ga}
\gamma^0=
\left(
\begin{array}{cc}
~I~ &~0~ \\
0 & -I
\end{array}
\right); \qquad
\gamma_5=
\left(
\begin{array}{cc}
~0~ &~ I~ \\~I~  &~ 0~
\end{array}
\right); \qquad
\gamma^i=
\left(
\begin{array}{cc}
~0~&\sigma^i\\-\sigma^i &~0~
\end{array}
\right).
\end{equation}
For the spinors at rest we have
\begin{equation}
\gamma_5 v_r=u_r \nonumber\, ,
\end{equation}
and then
\begin{eqnarray} \label{eq:Qu}
Q^\mu&=&\langle \bar u|\gamma^\mu-\gamma^\mu\gamma_5|v\rangle=\langle \bar u|\gamma^\mu-\gamma^\mu\gamma_5| \gamma_5 u\rangle = \langle \bar u|\gamma^\mu\gamma_5-\gamma^\mu|  u\rangle =-\langle \bar u|\gamma^\mu-\gamma^\mu\gamma_5| u\rangle \, .
\end{eqnarray}
Thus, apart from a global sign we can work with the $u$ spinors all the time.

Next we must care about how to combine the spins of the quark-antiquark to states of given angular momentum. Indeed, in the $ W q\bar{q}$ vertex of
Fig. \ref{fig:tau}, we shall have the matrix element
\begin{eqnarray}
ME=\langle m |~ Operator~ | m^{\prime} \rangle,
\end{eqnarray}
but we want to combine the spins to total angular momentum and for this we use for the antiparticles the rule of particle-hole conjugation \cite{bohr}, where the hole with $m^{\prime}$
behaves as a particle state according to
\begin{equation}
|hole, m^{\prime} \rangle \to (-1)^{\frac{1}{2}- m^{\prime}} |\frac{1}{2},-m^{\prime}\rangle \, .
\end{equation}
We can include the minus sign of Eq. \eqref{eq:Qu} and then we will implement the rule
\begin{equation}
|hole, m^{\prime} \rangle \to (-1)^{\frac{1}{2}+ m^{\prime}} |\frac{1}{2},-m^{\prime}\rangle \, .
\end{equation}
We shall, then, carry on the former phase and change the sign of  $m^{\prime}$ to combine spins in what follows.

The next step is to realize that for the spinors at rest and $\gamma^\mu$ matrices, Eqs. \eqref{eq:wfn} and Eq. \eqref{eq:ga},
$ \bar{u}\gamma^0 u \to  \langle \chi^{\prime} |\chi \rangle$,
\begin{equation}
\bar{u}\gamma^0 u \to  \langle \chi^{\prime} |\chi \rangle \nonumber \, ,
\end{equation}
which means, $\gamma^0$ becomes the operator $1$ with bispinors,
\begin{equation}
\bar{u}\gamma^i \gamma_5 u \to  \langle \chi^{\prime} | \sigma_i |\chi \rangle \nonumber \, ,
\end{equation}
hence, replacing $\gamma^i \gamma_5 $ by  $\sigma_i$, the Pauli matrices,  with bispinors.  The  rest of matrix elements are zero.
Then
\begin{eqnarray} \label{eq:Qu2}
Q_0&=& \langle \chi^{\prime} | 1 | \chi \rangle \equiv M_0  \nonumber \, , \\
Q_i&=&  \langle \chi^{\prime} | \sigma_i |\chi  \rangle \equiv N_i \, ,
\end{eqnarray}
Denoting for simplicity,
\begin{equation}\label{eq:new}
\overline{L}^{\mu\nu}= \overline{\sum} \sum  L^\mu {L^\nu}^\dagger  \, ,
\end{equation}
we can write

\begin{equation}\label{eq:L}
\begin{array}{ccc}
\overline{\sum} \sum  L^\mu {L^\nu}^\dagger Q_\mu Q_\nu^{\star}
&=\,~~~\overline{L}^{00} & \qquad M_0~ M^{\star}_0 \\
& + \,~~\overline{L}^{0i} &\qquad M_0 ~N^{\star}_i \\
& + \,~~~\overline{L}^{i0} &\qquad N_i ~M^{\star}_0 \\
&+ \,~~\overline{L}^{ij} & \qquad N_i ~N_j^{\star}  \, ,
\end{array}
\end{equation}
where in $M_0 M_0^{\star}$, $M_0 N_i^{\star}$, $N_i M_0^{\star}$, $N_i~ N_j^{\star}$ we shall sum over the final polarizations of the mesons produced.
$\overline{\sum} \sum  L^\mu {L^\nu}^\dagger $ is easily evaluated and we have
\begin{equation}\label{eq:LL}
\overline \sum  \sum  L^\mu {L^\nu}^\dagger = \frac{1}{m_{\nu} m_{\tau}}\left( p'^\mu p^\nu
 +p'^\nu p^\mu - g^{\mu\nu}p'\cdot p+i \epsilon^{\alpha\mu\beta\nu}p'_\alpha p_\beta\right),
\end{equation}
where we use the field normalization for fermions of Ref. \cite{mandl}.

Next we must evaluate $M_0$ and $N_i$ for the different $PP$, $PV$, $VP$ and $VV$ combinations. In order to implement the
hadronization of Fig. \ref{fig:tauh} we use  the $^3P_0$ model \cite{micu,oliver,close}, the essence of which is that
the $\bar{q}q$ introduced must  have parity $+$ and zero angular momentum.  Since $\bar{q}$ has negative parity  we need $L=1$
to restore parity, which forces the $\bar{q}q$ to couple to spin $S=1$ and then $L,S$ couple to $J=0$. We take the contribution of spin
\begin{equation}\label{eq:s3}
|1 S_3 \rangle =\sum_s {\cal C}(\frac{1}{2}\frac{1}{2}1;s, S_3-s) |\frac{1}{2},s\rangle  |\frac{1}{2},S_3-s \rangle \, ,
\end{equation}
where $|\frac{1}{2},S_3-s \rangle$ corresponds to the antiparticle $\bar{q}$ with sign and phase implicitly included and is considered as a normal particle state.
This is now coupled to $Y_{1,M_3}$ to give $J=0$.  Thus
\begin{eqnarray}\label{eq:00}
|00 \rangle &=&\sum_{M_3} {\cal C}(1 1 0; M_3,S_3)  Y_{1,M_3}({\hat{\bm{r}}})  |1,S_3 \rangle \nonumber\, ,\\
&=&\sum_{S_3} {\cal C}(1 1 0; -S_3,S_3)  Y_{1,-S_3}({\hat{\bm{r}}})  |1,S_3 \rangle \, .
\end{eqnarray}

 Next we must look at the spatial matrix element. For this we assume that for this low energy problem all the quark states are in their ground state.
 This assumption leads naturally to the weak chiral Lagrangians \cite{Gasser,Scherer}. Then we have
 \begin{equation}
ME({\bm{q}})=\int d^3{\bm{r}}\varphi_{\bar{u}}(r)\varphi_d(r)\varphi_q(r) \varphi_{\bar q}(r) e^{i {\bm q}\cdot {\bm r}} Y_{1,-S_3}({\hat{\bm{r}}})
\end{equation}
with ${\bm q}={\bm p_1}-{\bm p_2}$,  where $\bm{p_1}$, $\bm{p_2}$ are the momenta of the mesons produced. By means of
 \begin{equation}
 e^{i {\bm q}\cdot {\bm r}} = 4 \pi \sum_l i^l j_l (qr) \sum_{\mu}   Y_{l\mu}({\hat{\bm{q}}}) Y^*_{l\mu}({\hat{\bm{r}}})
\end{equation}
we obtain
 \begin{equation}
ME({\bm{q}})=i 4 \pi Y_{1,-S_3}({\hat{\bm{q}}}) \int r^2dr \varphi_{\bar{u}}(r)\varphi_d(r)\varphi_q(r) \varphi_{\bar q}(r) j_1 (qr) \, .
\end{equation}

As we have commented, we do not wish to evaluate this matrix element which involves large uncertainties,  but rather establish relationships between different decays
based exclusively on the flavor-spin structure. However, due to the fact that $j_1 (qr)$ go as $qr$, hence $q$, for low values of $q r$, and the fact that $q$
is very different for different decays, due to their different masses, the appropriate  procedure is to write
 \begin{equation}\label{eq:MEq}
ME({\bm{q}})=i 4 \pi q Y_{1,-S_3}({\hat{\bm{q}}}) \frac{1}{3}\int r^2dr  \prod_{i}\varphi_i(r) \frac{3j_1(qr)}{qr}r\equiv q Y_{1,-S_3}({\hat{\bm{q}}}) F(q) \, .
\end{equation}
where in the evaluation of $F(q)$ we  use the factor $\frac{3j_1 (qr)}{qr}$ in the integrand which goes to 1 as $q r \to 0$ and is a smooth function over
the range of $\prod_{i}\varphi_i(r)$. This allows for a better comparison of rates for different decays assuming  $F(q)$ the same for all of them since the quark
wave functions refer to the ground state in all cases that we study. This factor $q Y_{1,-S_3}({\hat{\bm{q}}})=\sqrt{\frac{3}{4\pi}}q_{-S_3} $ (in spherical basis) leads to the WPP coupling of chiral perturbation theory \cite{Gasser,Scherer}.

Once the integral over $d^3r$ is done and assuming $F(q)$ the same in all the decays,  the $|00 \rangle$ state of Eq. \eqref{eq:00} leads to
\begin{equation}\label{eq:00q}
|00 \rangle_q =\sum_{S_3} (-1)^{1+S_3}\frac{1}{\sqrt{3}} Y_{1,-S_3}({\hat{\bm{q}}})|1 S_3 \rangle \, ,
 \end{equation}
where we have permuted indices in ${\cal C}(1 1 0; -S_3,S_3)$ to obtain this Clebsch-Gordan coefficient (CGC) (we follow Rose conventions  and formulas for all the coming
Racah algebra \cite{rose})

  Next we must combine $|00 \rangle_q$  with the $d,\bar{u}$ spins to obtain the final $J M$,$J M^{\prime}$ angular momenta  of the two mesons produced. This is accomplished
  by means of the CGC and we have

\begin{eqnarray}
|J M \rangle &=&\sum_{m} {\cal C}( \frac{1}{2} \frac{1}{2} J; m, s, M) | \frac{1}{2} m \rangle |\frac{1}{2} s\rangle \nonumber\,  ,\\
|J' M' \rangle& =&\sum_{m'} {\cal C}( \frac{1}{2} \frac{1}{2} J'; S_3-s,-m', M') (-1)^{\frac{1}{2}+m'} | \frac{1}{2}, S_3-s \rangle |\frac{1}{2},-m'\rangle  \, ,
 \end{eqnarray}
which requires $m=M-s$,$m'=S_3-s-M'$,  and combined with $|00 \rangle_q$ of  Eqs. \eqref{eq:00q} and \eqref{eq:s3} lead to the matrix elements
\begin{eqnarray}\label{eq:ME}
 ME& =&-\frac{1}{\sqrt{3}} \sum_{S_3}\sum_{s} {\cal C}(\frac{1}{2} \frac{1}{2} 1; s, S_3-s, S_3)(-1)^{\frac{1}{2}-s-M'} q Y_{1,-S_3}({\hat{\bm{q}}})\, {\cal C}(\frac{1}{2} \frac{1}{2} J; M-s, s, M) \nonumber\,  \\
  &\times &  {\cal C}(\frac{1}{2} \frac{1}{2} J'; S_3-s,M'-S_3+s, M')
     \left\{
    \begin{array}{ll}
    \langle m |\,1\,| m' \rangle &~~~~~~~~ ($\rm{i}$)\, \\[2mm]
  \langle m | \sigma_i | m' \rangle &~~~~~~~~($\rm{ii}$)\,\\[2mm]
    \end{array}
   \right.
 \end{eqnarray}
\begin{itemize}
\item[(i)]  In the case of the operator $1$ leading to $M_0$ of Eq. \eqref{eq:Qu2} we have the extra constraint $m=m'=M-s$ and then $S_3=M+M'$.
\item[(ii)]  We shall work in spherical basis and evaluate  $\langle m | \sigma_\mu | m' \rangle (\mu=\pm 1,0)$. We have
$\langle m | \sigma_\mu | m' \rangle=\sqrt{3}\,{\cal C}(\frac{1}{2} 1 \frac{1}{2};m', \mu,m)$,  which induces the constraint
$m'+\mu=m$, $\mu=M-S_3+M'$.  We call $N_\mu$ the matrix element resulting from Eq. \eqref{eq:ME} in this case.
\end{itemize}

In Appendix A we evaluate these matrix elements explicitly for $PP$, $PV$, $VP$, $VV$ and we quote here the results.

\begin{itemize}
\item[(i)]  $M_0$
\begin{itemize}
\item[(a)] $PP$:  $J=0,J'=0$
\begin{equation}
M_0=0
\end{equation}
\item[(b)] $PV$:  $J=0,J'=1$
 \begin{equation}
  M_0=(-1)^{-M-M'}\frac{1}{\sqrt{6}} \,q\,   Y_{1,-(M+M')}({\hat{\bm{q}}})\, \delta_{M0}
\end{equation}

\item[(c)] $VP$:  $J=1,J'=0$  \\
 \begin{equation}
  M_0=(-1)^{-M-M'}\frac{1}{\sqrt{6}}  \,q\, Y_{1,-(M+M')}({\hat{\bm{q}}}) \,\delta_{M'0}
\end{equation}

\item[(d)] $VV$:  $J=1,J'=1$  \\
 \begin{equation}
  M_0=(-1)^{-M-M'}\frac{1}{\sqrt{3}}{\cal C}(1 1 1; M,M',M+M')  \,q \,Y_{1,-(M+M')}({\hat{\bm{q}}})
\end{equation}

\end{itemize}

\item[(ii)]$N_\mu$
\begin{itemize}
\item[(a)] $PP$:  $J=0,J'=0$
 \begin{equation}
N_\mu=\frac{1}{\sqrt{6}} \,q \, Y_{1,\mu}({\hat{\bm{q}}}) \,\delta_{M0} \, \delta_{M'0}
\end{equation}

\item[(b)] $PV$:  $J=0,J'=1$
 \begin{equation}
N_\mu=(-1)^{1-M'} \frac{1}{\sqrt{3}} \,q \, Y_{1,\mu-M'}({\hat{\bm{q}}})  {\cal C}(1 1 1; M',-\mu,M'-\mu) \, \delta_{M0}
\end{equation}

\item[(c)] $VP$:  $J=1,J'=0$  \\
 \begin{equation}
N_\mu=(-1)^{-M} \frac{1}{\sqrt{3}} \,q \, Y_{1,\mu-M}({\hat{\bm{q}}})   {\cal C}(1 1 1; M,-\mu,M-\mu) \, \delta_{M'0}
\end{equation}

\item[(d)] $VV$:  $J=1,J'=1$  \\
 \begin{eqnarray}\label{eq:nVV}
N_\mu &=& \frac{1}{\sqrt{6}} \,q \, Y_{1,\mu-M-M'}({\hat{\bm{q}}}) \{(-1)^{-M'} \delta_{\mu M}+ 2\,(-1)^{-M}{\cal C}(1 1 1; M,-\mu,M-\mu) \nonumber\,  \\
&\times  &  {\cal C}(1 1 1; M',-M-M'+\mu,-M+\mu) \}
 \end{eqnarray}
 \end{itemize}
\end{itemize}

The formulas obtained allow us to exploit selection rules for $G$-parity. Let us see how it proceeds. By inspecting the change when we permute
particle 1 and 2, taking into account that in this permutation $Y_{1,\nu}({\hat{\bm{q}}})=Y_{1,\nu}({\widehat{\bm{p_1-p_2}}})$ goes to
$Y_{1,\nu}({\widehat{\bm{p_2-p_1}}})=(-)^{1} Y_{1,\nu}({\widehat{\bm{p_1-p_2}}})$, we find the results of Table \ref{tab:sign}.

\begin{table}[h!]
\renewcommand\arraystretch{1.0}
\caption{ Signs resulting in the $M_0$, and $N_{\mu}$ amplitudes by permuting the order of the mesons. }
\centering
\begin{tabular}{ccccc }
\toprule[1.0pt]\toprule[1.0pt]
~~~~~&~~~~~ $PP$~~~~~  & ~~~~~ $PV$ ~~~~~& ~~~~~ $VP$~~~~~ &~~~~~  $VV$~~~~~ \\
\hline
$M_0$ & $0$  &  $-$ &  $-$ &  $+$ \\
\hline
$N_\mu$ &$ -$  &  $+$ &  $+$ &  $-$ \\
\midrule[1.0pt]\midrule[1.0pt]
\end{tabular}\label{tab:sign}
\end{table}

In the signs of Table \ref{tab:sign}  we have taken into account that when exchanging particle 1 and 2 in the $PV$ case we go to the $VP$ case. For the  case of
the $M_0$ amplitude there is no sign change (apart from $Y_{1,-(M-M')}({\hat{\bm{q}}}))$ in the formula to go from $ PV$  to  $VP$, but for the case of $VV$ we have
${\cal C}(1 1 1; M,M',M+M')=(-1)^{1+1-1} {\cal C}(1 1 1; M',M,M+M')$ and hence a change of sign.  On the other hand,  the situation  in the $N_\mu$ amplitude is opposite.
For $PP$ there is no change of sign, apart from  $Y_{1,\mu}({\hat{\bm{q}}})$. However, in the $PV$ to $ VP$ change we see a change of sign from the phase of CGC, apart from  $Y_{1,\mu-M}({\hat{\bm{q}}})$. Finally the case of $VV$ is more complicated but taking the $z$ axis such that  $\sigma_i$ becomes $\sigma_z$,  only $\mu=0$ contributes and one can explicitly see by inspection of all possible cases
 that the amplitude does not  change by exchanging the two particles, except for the $Y_{1,\mu-M-M'}({\hat{\bm{q}}})$.  Interestingly, in some cases the role of the first and second terms in Eq. \eqref{eq:nVV} are exchanged, but the sum remains the same.

Let us  use the result of Table \ref{tab:sign} to see the contribution of the channels shown in Eqs. \eqref{eq:PP21}, \eqref{eq:PV21}, \eqref{eq:VP21} and Eqs. \eqref{eq:PP31},
\eqref{eq:PV31},\eqref{eq:VP31},\eqref{eq:VV31}. If we take the $\pi^- \pi^0$ channel it comes with the combination $\pi^- \pi^0 - \pi^0 \pi^-$. As a consequence $N_\mu$ adds for the two terms
and we have a weight $ 2 \frac{1}{\sqrt{2}}$ for the $\pi^- \pi^0$ channel. On the other hand if we take $\pi^- \eta$,  $\pi^- \eta'$ they come with the combinations $\pi^- \eta + \eta \pi^- $,
$\pi^- \eta' + \eta' \pi^- $ and then the combination of the two terms cancel and we do not have $\pi^- \eta$, $\pi^- \eta'$  production. In the next subsection we shall see the relationship
of this to $G$-parity. We can proceed like that for the $\pi^- \rho^0$, $\rho^0 \pi^- $ there the two terms add in $M_0$ and cancel in $N_\mu$. The opposite happens to the $\pi^- \omega$ channel
and so on. A consequence of that, although  there is no $G$-parity in this case, is that the terms $K^- \eta$, $\eta K^-$ also  add in $N_\mu$ to give a weight of $\frac{2}{\sqrt{3}}$ for the $K^- \eta$ channel,
and $K^- \eta'$,   $\eta' K^-$ also lead to a weight  $-\frac{1}{\sqrt{6}}$  for $N_\mu$ for the $K^- \eta'$ channel. For the same reasons the contribution of $\eta  K^{*-}$, $K^{*-} \eta$  lead to a
weight $-\frac{2}{\sqrt{3}}$ in $M_0$ for $\eta  K^{*-}$ and zero in $N_\mu$, while $\eta'  K^{*-}$ and $ K^{*-}\eta'$ combine to give a weight $\frac{1}{\sqrt{6}}$ in  $M_0$ and  $\frac{3}{\sqrt{6}}$
in $N_\mu$ for  $\eta'  K^{*-}$ . Altogether we find the weight of  $M_0$,  $h_i$, and $N_\mu$, $\overline{h}_i$, for the different channels in Table \ref{tab:h1}.
Since we  want to evaluate ratios, the Cabibbo suppressed modes go with $\frac{\sin\theta_c}{\cos\theta_c}=\tan\theta_c$ with respect to the allowed modes, with $\theta_c$
the Cabibbo angle, $\cos\theta_c=0.97427$.

\begin{table}[h]
\renewcommand\arraystretch{0.92}
\caption{ Weight for the different channels after taking into account the $M_1 M_2$ and  $M_2 M_1$  components as they appear in the hadronization. }
\centering
\begin{tabular}{cccc ccc  }
\toprule[1.0pt]\toprule[1.0pt]
Channels &~ $h_i$ (for $M_0$) ~ & ~$\overline{h}_i$ (for $N_\mu$)~ &{~~~~~~~~~~}& ~Channels~~&~~ $h_i$ (for $M_0$)  & ~$\overline{h}_i$ (for $N_\mu$) \\
\hline
$\pi^{0}  \pi^{-}$  & $0$ &  $\sqrt{2}$ &&  $K^{*0} K^{*-} $ & $1$ &  $1$   \\
$\pi^{-}  \eta$  & $0$ &  $0$            && $K^{-} \pi^{0} $  & $0$ &  $\frac{1}{\sqrt{2}}\tan\theta_c$ \\
$\pi^{-}  \eta^{\prime}$  & $0$ &  $0$  &&  $\bar{K}^{0} \pi^{-} $  & $0$ & $\tan\theta_c$ \\
$\pi^{-} \rho^{0}$  & $\sqrt{2}$ &  $0$ &&  $K^{-} \rho^{0} $  &  $\frac{1}{\sqrt{2}}\tan\theta_c$ &  $\frac{1}{\sqrt{2}}\tan\theta_c$ \\
$\pi^{-} \omega$  & $0$  &   $\sqrt{2}$  &&  $K^{-} \omega $  &  $\frac{1}{\sqrt{2}}\tan\theta_c$ &  $\frac{1}{\sqrt{2}}\tan\theta_c$\\
$\pi^{0} \rho^{-}$  & $-\sqrt{2}$ &  $0$ &&  $\bar{K}^{0} \rho^{-} $  & $\tan\theta_c$ & $\tan\theta_c$ \\
$\eta \rho^{-}$  & $0$  & $ \frac{2}{\sqrt{3}}$  &&  $\eta K^{*-}$  &  $ -\frac{2}{\sqrt{3}}\tan\theta_c$ &  $0$ \\
$\eta^{\prime} \rho^{-}$  & $0$  & $ \frac{2}{\sqrt{6}}$  && $\eta^{\prime} K^{*-}$  &  $ \frac{1}{\sqrt{6}}\tan\theta_c$ &  $ \frac{3}{\sqrt{6}}\tan\theta_c$\\
$\rho^{-} \rho^{0}$  & $0$ &  $\sqrt{2}$  &&  $K^{*-} \pi^{0} $  &  $ \frac{1}{\sqrt{2}}\tan\theta_c$ &  $ \frac{1}{\sqrt{2}}\tan\theta_c$ \\
$\rho^{-} \omega$  & $\sqrt{2}$&  $0$    &&  $K^{*0} \pi^{-} $  &  $\tan\theta_c$ &  $ \tan\theta_c$ \\
$K^{-}  \eta$  & $0$ &  $ \frac{2}{\sqrt{3}}\tan\theta_c$   &&  $\phi  K^{-} $  &  $\tan\theta_c$ &  $ \tan\theta_c$ \\
$K^{-} \eta^{\prime}$  & $0$ & $ -\frac{1}{\sqrt{6}}\tan\theta_c$  && $K^{*-} \rho^{0} $  &  $ \frac{1}{\sqrt{2}}\tan\theta_c$ &  $ \frac{1}{\sqrt{2}}\tan\theta_c$\\
$K^{0} K^{-}$  & $0$ &  $1$   && $K^{*-} \omega $  &  $ \frac{1}{\sqrt{2}}\tan\theta_c$ &  $ \frac{1}{\sqrt{2}}\tan\theta_c$\\
$K^{0} K^{*-}$  & $1$ &  $1$ &&  $\bar{K}^{*0} \rho^{-} $  &  $\tan\theta_c$ &  $ \tan\theta_c$\\
$K^{*0} K^{-} $  & $1$ &  $1$ && $\phi  K^{*-} $  &  $\tan\theta_c$ &  $ \tan\theta_c$\\
\midrule[1.0pt]\midrule[1.0pt]
\end{tabular}\label{tab:h1}
\end{table}

\subsection{$G$-parity considerations }
\label{sec:Gparity}
Taking into account the $G$-parity of the mesons, $\pi (-)$, $\eta (+)$, $\eta' (+)$, $\rho (+)$, $\omega (-)$, $\phi (-)$ we can
associate a $G$-parity  to all nonstrange $M_1M_2$ pairs.  On the other hand, the $G$-parity can already be established from the original $d\bar{u}$ pair and the operator producing them,
$1$ or $\sigma_i$.  We know that the $G$-parity for quarks belonging to the same isospin  multiplet is given  by \cite{close}
\begin{equation}
G=(-1)^{L+S+I} \, ,
\end{equation}
but here $L=0, I=1$ and $S=0$ for the $1$ operator and $S=1$ for the $\sigma_i$ operator. Thus we have $G$-parity negative for the $1$ operator and positive parity for the $\sigma_i$ operator.
As a consequence we find the result of   Table  \ref{tab:sign2} for the different channels.
\begin{table}[h]
\renewcommand\arraystretch{1.1}
\caption{ Contributions of the different non-strange $M_1 M_2$ pairs. The cross indicates non zero  contribution. }
\centering
\begin{tabular}{cccc }
\toprule[1.0pt]\toprule[1.0pt]
~~~~~ Channels~~~~~ &~~~~G-parity~~~~  & ~~~~~$M_0$  ~~~~~& ~~~~~$N_{\mu}$ ~~~~~  \\
\hline
$\pi^{-}  \pi^{0}$ & $+$  &  $0$ &  $\times$ \\
$\pi^{-} \eta$ & $-$  &  $0$ &  $0$ \\
$\pi^{-} \eta^{\prime}$ & $-$  &  $0$ &  $0$ \\
$\pi^{-} \rho^{0}$ & $-$  & $\times$ &  $0$ \\
$\pi^{-} \omega$ & $+$  &  $0$  &  $\times$ \\
$\pi^{0} \rho^{-}$ & $-$  & $\times$ &  $0$ \\
$\eta \rho^{-}$ & $+$  &  $0$ & $\times$ \\
$\eta^{\prime} \rho^{-}$ & $+$  &  $0$ & $\times$ \\
$\rho^{-} \rho^{0}$ & $+$  &  $0$ & $\times$ \\
$\rho^{-} \omega$ & $-$  &  $\times$ &  $0$   \\
\midrule[1.0pt]\midrule[1.0pt]
\end{tabular}
\label{tab:sign2}
\end{table}

We can see comparing with Table  \ref{tab:h1} that the $G$-parity rules of  Table  \ref{tab:sign2} coincide with what we obtained in   Table \ref{tab:h1} considering the order of the $M_1M_2$ pairs
in the hadronization and the explicit formulas for $M_0$ and $N_\mu$, with their properties under the exchange of $M_1$ and $M_2$.
We  can see that the matrix elements are all zero for $\pi^-\eta$, $\pi^-\eta'$ cases, which shows from
a different perspective that it is  the value of $M_0=0$ for $PP$ and $G$-parity what makes the matrix elements zero, in coincidence with results obtained through
different methods \cite{Leroy,Escribano,roig}.
Note, however, that the $G$-parity  restrictions have clear repercussions on which of the $M_0$  or  $N_\mu$ terms contribute to the process.

\section{Evaluation of $\overline{\sum} \sum \left|t\right|^2$ for the different processes}
Following the  nomenclature adopted in Eq. \eqref{eq:new} we  must evaluate
\begin{equation}\label{eq:dai}
\overline{\sum} \sum \left|t\right|^2=  \overline{L}^{00} M_0~ M^{\star}_0+\overline{L}^{0i}   M_0 ~N^{\star}_i +\overline{L}^{i0}   N_i ~M_0^{\star} + \overline{L}^{ij}  N_i ~N_j^{\star}
\end{equation}
and in this equation we must sum over $M,M'$ the spin third components of $J,J'$.  This is done in Appendix B and here we summarize the results.

\begin{itemize}
\item[1)]  $PP, J=0, J'=0$\\
Only the term $N_\mu$ contributes and we obtain
\begin{equation}\label{eq:ft1}
\overline{\sum} \sum \left|t\right|^2= \frac{1}{m_\tau m_\nu} \frac{1}{2 \pi}\, {\widetilde p^2_1}\, \left(E_\tau E_\nu -\frac{{\bm{p}}^2 }{3} \right) \, ,
\end{equation}
which, as discussed previously is evaluated in the frame where the system $M_1,M_2$ is at rest, $p$ is the momentum of the $\tau$, or $\nu$, in that frame, given by
\begin{equation}\label{eq:newlabel}
p=\frac{\lambda^{1/2}(m^2_\tau,m^2_\nu,M^2_{\rm inv}(M_1 M_2))}{2 M_{\rm inv}(M_1 M_2)}\, ,
\end{equation}
$E_\tau=\sqrt{m_\tau^2+p^2}$, $E_\nu=p$ and $\overline{L}^{\mu\nu}$ of Eq. \eqref{eq:LL} is evaluated in this frame too. In  Eq. \eqref{eq:ft1}
${\widetilde p_1}$ is the momentum of the meson $M_1$ in the same frame where the system $M_1 M_2$ is at rest,
\begin{equation} \label{eq:new2}
\widetilde{p}_1=\frac{\lambda^{1/2}(M^2_{\rm inv}(M_1 M_2), m_{M_1}^2, m_{M_2}^2)}{2 M_{\rm inv}(M_1 M_2)}\, .
\end{equation}

\item[2)]  $PV, J=0, J'=1; VP, J=1, J'=0 $
\begin{itemize}
\item[a)] The $\overline{L}^{00} M_0~ M^{\star}_0$ contribution, summed over $M, M'$ gives
\begin{equation}\label{eq:ft2}
\overline{\sum} \sum \left|t_a\right|^2= \frac{1}{m_\tau m_\nu} \frac{1}{2 \pi} \,{\widetilde p^2_1} \,\left(E_\tau E_\nu + {\bm{p}}^2 \right) \, .
\end{equation}
\item[b)] The $M_0 ~N^{\star}_i$ and $N_i ~M_0^{\star}$ combinations gives zero.
\item[c)] The $ N_i ~N_j^{\star}$ term of Eq. \eqref{eq:dai} gives
\begin{equation}\label{eq:ft3}
\overline{\sum} \sum \left|t_c\right|^2= \frac{1}{m_\tau m_\nu} \frac{1}{\pi}\, {\widetilde p^2_1} \,\left(E_\tau E_\nu -\frac{{\bm{p}}^2 }{3} \right)\, .
\end{equation}
\end{itemize}
\item[3)]  $VV, J=1, J'=1$
\begin{itemize}
\item[(a)]  The $\overline{L}^{00} M_0~ M^{\star}_0$ term gives
\begin{equation}\label{eq:ft4}
\overline{\sum} \sum \left|t_a\right|^2= \frac{1}{m_\tau m_\nu} \frac{1}{\pi} \,{\widetilde p^2_1}\, \left(E_\tau E_\nu + {\bm{p}}^2 \right) \, .
\end{equation}
\item[(b)]  The $\overline{L}^{0i} M_0~ N^{\star}_i$,  $\overline{L}^{i0} N_i~ M^{\star}_0$ terms gives zero.
\item[(c)]  The $\overline{L}^{ij} N_i~ N^{\star}_j$  term gives the result
\begin{equation}\label{eq:ft5}
\overline{\sum} \sum \left|t_c\right|^2= \frac{1}{m_\tau m_\nu} \frac{1}{\pi} \,{\widetilde p^2_1}\, \frac{7}{2} \left(E_\tau E_\nu - \frac{{\bm{p}}^2 }{3}\right) \, .
\end{equation}
\end{itemize}
\end{itemize}

Taking into account the weights $h_i$, $\overline{h}_i$ of Table \ref{tab:h1}, we get finally the following result

\begin{itemize}
\item[1)]  $PP, J=0, J'=0$\\
\begin{equation}\label{eq:ff1}
\overline{\sum} \sum \left|t\right|^2= \overline{h}^2_i  \frac{1}{m_\tau m_\nu} \frac{1}{2 \pi}\, {\widetilde p^2_1}\, \left(E_\tau E_\nu -\frac{{\bm{p}}^2 }{3} \right) \, .
\end{equation}
\item[2)]  $PV, J=0, J'=1; VP, J=1, J'=0 $
\begin{equation}\label{eq:ff2}
\overline{\sum} \sum \left|t\right|^2=  \frac{1}{m_\tau m_\nu} \frac{1}{2 \pi}\, {\widetilde p^2_1}\,
 \left[{h}^2_i \left(E_\tau E_\nu + {\bm{p}}^2 \right) + 2\,\overline{h}^2_i \left(E_\tau E_\nu -\frac{ {\bm{p}}^2 }{3}\right) \right] \, .
\end{equation}
\item[3)]  $VV, J=1, J'=1$
\begin{equation}\label{eq:ff3}
\overline{\sum} \sum \left|t\right|^2=  \frac{1}{m_\tau m_\nu} \frac{1}{ \pi}\, {\widetilde p^2_1}\,
 \left[{h}^2_i \left(E_\tau E_\nu + {\bm{p}}^2 \right) + \frac{7}{2}\,\overline{h}^2_i \left(E_\tau E_\nu -\frac{{\bm{p}}^2 }{3} \right) \right]\, .
\end{equation}
\end{itemize}

In the former equations the angle integrations are already  done in a way that finally we must take into account the full phase space with the angle independent expressions
obtained in the former equations and we obtain
\begin{equation}\label{eq:dGdM}
\frac{ d\Gamma}{dM_{\rm inv}(M_1 M_2)} =  \frac{2\,m_\tau 2\, m_\nu}{(2\pi)^3} \frac{1}{4 m^2_\tau}\, p_\nu {\widetilde p_1}\, \overline{\sum} \sum \left|t\right|^2 \,,
\end{equation}
where $p_\nu$ is the neutrino momentum in the $\tau$ rest frame
\begin{equation}
p_\nu=\frac{\lambda^{1/2}(m^2_\tau,m^2_\nu,M^2_{\rm inv}(M_1 M_2))}{2 M_\tau}\, ,
\end{equation}
and  $ {\widetilde p_1}$ the momentum of $M_1$  in the $M_1, M_2$ rest frame  given in  Eq. \eqref{eq:new2}.
The mass distribution of Eq. \eqref{eq:dGdM} is then integrated over the $M_1 M_2$ invariant mass in order to obtain the width.

\begin{table}[h]
\renewcommand\arraystretch{1.1}
\caption{ $h_i$ and $\overline{h}_i$ coefficient for  different channels with the two final mesons in $s$-wave. }
\centering
\begin{tabular}{ccc }
\toprule[1.0pt]\toprule[1.0pt]
~~~~&~~~~~  for $M_0$~~~~~  & ~~~~~ for $N_\mu$ ~~~~~ \\
~channels~~~~&~~~~~ $h_i$ ~~~~~  & ~~~~~$\overline{h}_i$ ~~~~~ \\
\hline
$\pi^{-} \rho^{0}$  &  $0$ & $\sqrt{2}$  \\
$\pi^{-} \omega$  &   $\sqrt{2}$ & $0$  \\
$\pi^{0} \rho^{-}$   &  $0$ & -$\sqrt{2}$\\
$\eta \rho^{-}$ & $ \frac{2}{\sqrt{3}}$   & $0$   \\
$\eta^{\prime} \rho^{-}$   & $ \frac{2}{\sqrt{6}}$ & $0$   \\
$\rho^{-} \rho^{0}$ &  $\sqrt{2}$   & $0$  \\
$\rho^{-} \omega$  &  $0$ & $\sqrt{2}$   \\
$\eta K^{*-}$  &   $0$&  $ -\frac{2}{\sqrt{3}}\tan\theta_c$ \\
$\eta^{\prime} K^{*-}$  &  $ \frac{3}{\sqrt{6}}\tan\theta_c$ &  $ \frac{1}{\sqrt{6}}\tan\theta_c$ \\
\midrule[1.0pt]\midrule[1.0pt]
\end{tabular}
\label{tab:hs}
\end{table}

\section{$S$-wave decays}
In the previous sections we have assumed that the quarks $d,\bar{u}$  of  Fig. \ref{fig:tau} are produced in their ground state, this leads to a negative parity $q\bar{q}$ state,
which makes the pair of  mesons after  the hadronization to be produced in $p$-wave and this is in agreement  with the results of chiral perturbation theory for $\tau^-$ decay
into  $\nu_\tau$ and  a pair of pseudoscalar mesons.

We shall extrapolate the scheme to  pseudoscalar-vector and vector-vector production, but we can anticipate that, since  the masses of these mesons are larger, the resulting momenta for
the mesons are much smaller and the $p$-wave mechanism will lead to very small widths. Certainly,  in this case, $s$-wave production shall be  preferable. There is just one inconvenience. Two mesons with
negative parity and $s$-wave  have positive parity. This means that the $d\bar{u}$  must be produced in an $L'=1$ state. This is accomplished
creating one quark in $L'=1$ state.

The formalism in this case proceed in a total analogy to what we have done before. There is only one difference. Since an $Y(L',M_3',1)$ is introduced, we have now two
spherical harmonics.  This one and the one from the $^3P_0$ model,  and they must combine to final  $s$-wave. Hence
\begin{equation}
Y_{1 M_3} Y_{1 M'_3}=\sum_{l} \left[\frac{3\cdot3}{4\pi(2l+1)} \right] {\cal C}(1 1 l; M_3,M'_3){\cal C}(1 1 l;0,0,0) Y_{l,M_3+M_3'}
\end{equation}
which can have $l=0,2$ for parity reasons and we then choose $l=0$. Evaluating explicitly the CGC we obtain
\begin{equation}
Y_{1 M_3} Y_{1 M'_3} \to \frac{1}{4\pi} (-1)^{M_3} \delta_{M_3,-M'_3}
\end{equation}
And the rest of calculations proceed as in the case of $p$-wave, only the
$Y_{1,\mu}({\bf\hat{q}})$ does not appear.  Also the form factor now  implies $j_0 (qr)$ instead  of $j_1 (qr)$  and there is no need to take the
 factor $q$ outside the integral. We obtain the results:
\begin{itemize}
\item[1)] $M_0$
\begin{itemize}
\item[a)]  $PP, J=0, J'=0$
\begin{equation}\label{eq:st1}
M_0=0
\end{equation}
\item[b)]  $PV, J=0, J'=1$\\
\begin{equation}\label{eq:st2}
M_0=\frac{1}{\sqrt{6}}\frac{1}{4\pi}
\end{equation}
\item[b)]  $VP, J=1, J'=0$\\
\begin{equation}\label{eq:st3}
M_0=\frac{1}{\sqrt{6}}\frac{1}{4\pi}
\end{equation}
\item[d)]  $VV, J=1, J'=1$\\
\begin{equation}\label{eq:st4}
M_0=\frac{1}{\sqrt{3}}\frac{1}{4\pi} {\cal C}(1 1 1; M,M',M+M')
\end{equation}
\end{itemize}
\item[2)] $N_\mu$
\begin{itemize}
\item[a)]  $PP, J=0, J'=0$
\begin{equation}\label{eq:st5}
N_\mu=\frac{1}{\sqrt{6}}\frac{1}{4\pi} \,\delta_{M0}\,\delta_{M'0}\,(-1)^{-\mu}
\end{equation}
\item[b)]  $PV, J=0, J'=1$
\begin{equation}\label{eq:st6}
N_\mu= -(-1)^{-\mu} \frac{1}{\sqrt{3}}\frac{1}{4\pi} {\cal C}(1 1 1; M',-\mu,M'-\mu)\,\delta_{M0}
\end{equation}
\item[c)]  $VP, J=1, J'=0$
\begin{equation}\label{eq:st7}
N_\mu= (-1)^{-\mu} \frac{1}{\sqrt{3}}\frac{1}{4\pi} {\cal C}(1 1 1; M,-\mu,M-\mu)\,\delta_{M'0}
\end{equation}
\item[d)]  $VV, J=1, J'=1$
\begin{equation*}
N_{\mu}= \frac{1}{\sqrt{6}} \frac{1}{4\pi}  \Big\{ \delta_{M \mu} + 2 \, (-1)^{^{-\mu}-M'} \ \mathcal{C}\left( 1 1 1; M, -\mu, M-\mu \right)
\end{equation*}
\begin{equation}
\label{eq:8-1}
 \times \mathcal{C}\left( 1 1 1; M', -M-M'+\mu, -M+\mu \right) \Big\}
\end{equation}
\end{itemize}
\end{itemize}

In this case table \ref{tab:sign} is changed and under the exchange of the two mesons we obtain opposite signs  than in this table because
we do not have the $Y_{1,\mu}({\bf\hat{q}})$ factor. As a consequence the weights of some channels, particularly those of defined $G$-parity
are changed. Note that now the rule $(-1)^{L+S+I}$  for the $G$-parity implies positive $G$-parity for the operator "1" and negative $G$-parity for the
operator $\sigma_i$. As a consequence we get the results of table \ref{tab:hs} for the new weights of the channels involved. The rest do not change.
The final formulas for  $\overline{\sum} \sum \left|t\right|^2$, up to a global normalization,  are  the same for $p$-wave removing
the factor $\widetilde{p}_1^2$, concretly:

\begin{itemize}
\item[1)]  $PP, J=0, J'=0$\\
\begin{equation}\label{eq:ff1}
\overline{\sum} \sum \left|t\right|^2=   \frac{1}{m_\tau m_\nu}  \left(\frac{1}{4\pi}\right)^2 \left(E_\tau E_\nu -\frac{{\bm{p}}^2 }{3} \right)  \frac{1}{2}  \overline{h}^2_i
\end{equation}
\item[2)]  $PV, J=0, J'=1; VP, J=1, J'=0 $
\begin{equation}\label{eq:ff2}
\overline{\sum} \sum \left|t\right|^2=  \frac{1}{m_\tau m_\nu} \left(\frac{1}{4\pi}\right)^2
 \left[\left(E_\tau E_\nu + {\bm{p}}^2 \right) \frac{1}{2}{h}^2_i   +  \left(E_\tau E_\nu -\frac{ {\bm{p}}^2 }{3}\right) \overline{h}^2_i \right]
\end{equation}
\item[3)]  $VV, J=1, J'=1$
\begin{equation}\label{eq:ff3}
\overline{\sum} \sum \left|t\right|^2=  \frac{1}{m_\tau m_\nu}\left(\frac{1}{4\pi}\right)^2
 \left[\left(E_\tau E_\nu + {\bm{p}}^2 \right) {h}^2_i   + \frac{7}{2} \left(E_\tau E_\nu -\frac{ {\bm{p}}^2 }{3}\right) \overline{h}^2_i \right]
\end{equation}
\end{itemize}

\section{Results}

In Table \ref{tab:p1} we show the results for the decays in Table  \ref{tab:h1} assuming the mesons are in $p$-wave. We should be careful selecting the data because in some cases a strong resonance can appear. This is the case of $\tau^- \to \pi^{0}  \pi^{-} \nu_{\tau}  $  where the $\rho^-(770)$ can be formed and decay to $\pi^{0}  \pi^{-} $. We should note that the $\tau^- \to  \nu_{\tau}\rho^{-}$  decay  does not require the hadronization since a $q\bar{q}$ can already produce the  $\rho^-$ \cite{Pelaez:2015qba}. In this case the rate of $\rho^-$ production should be bigger than the nonresonance
$\pi^{0}  \pi^{-} \nu_{\tau} $ which is actually the case experimentally.  We calculate only the non resonance part of the decay, which involves the hadronization and we compare  with the "non resonance"  results of the PDG \cite{pdg}.  The same can be said about the $\pi^{-}  \bar{K}^{0} \nu_{\tau}$ and $\pi^{0} K^{-} \nu_{\tau}$.
 In fact, for $\pi^{-}  \bar{K}^{0} \nu_{\tau}$  the whole branching  ratio  is $8.4  \times 10^{-3}$  while the "non resonance" part is $5.4  \times 10^{-4}$ .
 In this case the resonance part comes from $\tau^- \to  K^{*-} \nu_{\tau}$.   For the   $\pi^{0} K^{-} \nu_{\tau}$ the PDG only quotes the whole branching  ratio.  We have estimated  the  non resonance  part as explained in the footnote of Table \ref{tab:p1}.

 If we look at the first block of  Table \ref{tab:p1} for decay to two pseudoscalars, we find that fixing our normalization to $K^{-} K^{0} \nu_{\tau}$ the rates obtained  in the other cases are close  to experiment within a factor of two or less. The rates obtained  for  $\eta \pi^{-} \nu_{\tau}$  and $\eta' \pi^{-} \nu_{\tau}$  are zero in our case, and experimentally the upper bounds are very  small.  For the case of  $\eta' K^{-} \nu_{\tau}$  we also get a value of the branching  ratio which is smaller than the experimental upper bound. The exception to the rule is the $\tau^- \to \pi^{0} \pi^{-} \nu_{\tau}$  that in our case is about one order of magnitude  bigger than experiment. This already indicates that the form factor of Eq. \eqref{eq:MEq}, with $q$ quite big and   $\frac{3j_1 (qr)}{qr}$ in the integrand, which we have assumed equal for all decays, should be smaller in the  case of $\pi^{0} \pi^{-} \nu_{\tau}$ production.  We should also note that we are taking a pion as a simple $q\bar{q}$, but this light Goldstone boson should be more complicated. Our results, and the discrepancies found, could serve as a tool of comparison  for theoretical models of this form factor.
As to the second block in Table \ref{tab:p1}, for $PV$, and $VV$ decay, what we observe is that the assumption of $p$-wave in the mesons leads systematically  to very small results compared to the experiment. There are two cases where  the discrepancies are larger  than in the other cases.  This occurs for  $\tau^- \to  K^{-} \rho^{0} \nu_{\tau} $ and $\tau^- \to  \bar{K}^{0}\rho^{-}  \nu_{\tau}$.  This has to be understood as a large contribution from the resonance  $K_1(1270)$ decaying into $\bar{K} \rho$, as found in \cite{Asner}, while we only calculate the  non resonance contributions. Yet, the findings of that work  are illustrative because the $K_1(1270)$  couples to $\bar{K} \rho$ in $s$-wave \cite{lutz,rocaSingh,geng}, which clearly indicate that $PV$ and $VV$  proceed  via $s$-wave meson-meson production, not $p$-wave.
We also take into account the mass distributions for the particles that have a width, but this leads to effects of the order of $10-20\%$ for the cases where are data, and do not improve the large discrepancies  found.

As mentioned, the experimental data for $\tau^- \to PV $  or $\tau^- \to VV $ indicate that $p$-wave is not adequate and instead the decays proceeds with the two mesons  in  $s$-wave. In Tables \ref{tab:s1}, \ref{tab:s2}, we show the results for the $\tau^- \to PV \nu_{\tau} $  or $\tau^- \to VV \nu_{\tau}$  with and without the convolution to take account the mass distribution of the vector mesons that have a width. This has been done according to the following  formulas. In the case of only one vector we make the convolution
\begin{equation}
\Gamma_{\tau\to\nu_\tau M_1 M_2}=\frac{1}{N} \int_{(M_1-2\Gamma_1)^2}^{(M_1 + 2\Gamma_1)^2} dm_1^2 \left(-\frac{1}{\pi}\right)  {\rm Im}  D_{M_1}(m_1)
\int dM_{\rm inv} \frac{d\Gamma(m_1,m_2)}{dM_{\rm inv} (12)} \,,
\label{eq:convol}
\end{equation}
where $D(m_1)$ is the vector propagator,
\begin{equation}
D(m_1)=\frac{1}{m_1^2-m_R^2+ i \,\Gamma_R  \, m_R}  \,.
\end{equation}
and $N$ is the normalization factor
\begin{equation}
N= \int_{(M_1-2\Gamma_1)^2}^{(M_1 + 2\Gamma_1)^2} dm_1^2 \left(-\frac{1}{\pi}\right)  {\rm Im}  D_{M_1}(m_1) \, .
\end{equation}
For the case of two vectors we make a double convolution as
\begin{eqnarray}
\Gamma_{\tau\to\nu_\tau M_1 M_2}&=&\frac{1}{N'} \int_{(M_1-2\Gamma_1)^2}^{(M_1 + 2\Gamma_1)^2} dm_1^2 \left(-\frac{1}{\pi}\right)  {\rm Im}  D_{M_1}(m_1) \nonumber\, \\
&\times&\int_{(M_2-2\Gamma_2)^2}^{(M_2 + 2\Gamma_2)^2} dm_2^2 \left(-\frac{1}{\pi}\right)  {\rm Im} D_{M_2}(m_2)
\int dM_{\rm inv} \frac{d\Gamma(m_1,m_2)}{dM_{\rm inv} (12)}  \,,
\label{eq:convo2}
\end{eqnarray}
where \begin{equation}
N'= \int_{(M_1-2\Gamma_1)^2}^{(M_1 + 2\Gamma_1)^2} dm_1^2 \left(-\frac{1}{\pi}\right)  {\rm Im} D_{M_1}(m_1)  \int_{(M_2-2\Gamma_2)^2}^{(M_2 + 2\Gamma_2)^2} dm_2^2 \left(-\frac{1}{\pi}\right)  Im D_{M_2}(m_2)\, .
\end{equation}

When performing the convolution, some of the decays forbidden in Table \ref{tab:s1}, as $\tau^- \to \eta^{\prime}  K^{*-}\nu_{\tau} $ and
$\tau^- \to K^{*0}  K^{*-}\nu_{\tau}  $, are now allowed,  and finite results arise in Table \ref{tab:s2}, although with very small rates. By looking at Table \ref{tab:s2}
and normalizing the results to the  $\tau^- \to \eta K^{*-} \nu_{\tau}$ branching ratio, we obtain fair results compared to experiment within a factor of about two, with
two exceptions:   $\tau^- \to  K^{-} \rho^{0} \nu_{\tau}  $  and $\tau^- \to  \bar{K}^{0}\rho^{-}  \nu_{\tau}$. As discussed previously, these two decays have a large contribution from the $K_1(1270)$  resonance  \cite{Asner} and thus, with the non resonance part that we calculate we underestimate the experimental results  by
about a factor three or more. This can be used in an opposite direction: a gross underestimation of the rates that we have calculated compared with future experiments
would be indicative of substantial resonance contribution, which can stimulate the research for such resonance  in the mass distribution.

It is also worth mentioning that in the work of \cite{rocaSingh,geng} this  $K_1(1270)$ resonance  were found coupling mostly to $\pi K^*$ and $K\rho$. The fair agreement
with the data of  $\tau^- \to \bar{K}^{*0} \pi^{-}  \nu_{\tau}  $ should be looked with caution, because we expect some overestimation due to the light pion mass, which indicates that there is  room of a resonance contribution, in this case  one of the two  $K_1(1270)$. Something similar could be said about the
$\tau^- \to \pi^{-} \rho^{0} \nu_{\tau}$ and  $\tau^- \to \pi^{0} \rho^{-} \nu_{\tau}$ decays. We should also expect an overestimation due to the small pion mass but we instead
underestimate the data  by about a factor of two. This again has to be looked with the perspective that the $\pi\rho$ couples strongly to the $h_1(1170)$ and
$a_1(1260)$ resonanes\cite{rocaSingh}.

For vector-vector there is also work leading to  dynamically generated resonance from the $VV$ interaction \cite{molina,gengvector}. However we do not have data for $\tau^-$ decay into $\nu_{\tau}$ and $VV$, something that could change in the future. In that case the comparison  of the measured decay rates  with our predictions would be of interest.

Finally, we should also mention that the formalism discussed here can be considered as a starting point to study the final state interaction of $M_1 M_2$,
eventually leading to dynamically generated resonances.  It would be most interesting to study experimently in detail invariant mass distributions in the
$\tau^- \to  M_1 M_2 \nu_{\tau}$  decays. One  case that has deserved much attention  in the  $\tau^- \to \pi \rho \nu_{\tau}$ via the    $a_1(1260)$ \cite{a1exp}.
In \cite{Leupold}  this decay  is done via  $\tau^- \to \nu_{\tau} PV$ , with $PV$ coupled channels that generate the   $a_1(1260)$, which decays into $\pi\rho$.
A different perspective, from the  point of view of resonance effective theory, including explicitly the  $a_1(1260)$ resonance, is given in \cite{isgur89,Pich}. A high precision   is obtained in the data and  one can think that such precision could be reached in other decays.
In the approach of \cite{Leupold} one would take the amplitudes evaluated here for $\tau^- \to  \nu_{\tau} \widetilde{M}_1 \widetilde{M}_2 $  with all possible coupled channels that lead to a given resonance, then  propagate $\widetilde{M}_1 \widetilde{M}_2 $  as they would do in scattering theory,  and later these
$\widetilde{M}_1 \widetilde{M}_2$ mesons are coupled to $M_1 M_2$,
which are the observed  mesons. The transition  of $\widetilde{M}_1 \widetilde{M}_2$ to $M_1 M_2$  is given by the $MM \to MM$ matrix that contains
information  on the resonance \cite{rocaSingh,gengvector}.

\begin{table}[h!]
\renewcommand\arraystretch{0.84}
\caption{Branching ratios for $\tau^- \to \nu_{\tau} M_1 M_2$  in $p$-wave normalized by $\tau^- \to K^{-} K^{0} \nu_{\tau}  $.  }
\centering
\begin{tabular}{l  c  c  c }
\toprule[1.0pt]\toprule[1.0pt]
{Decay process~} ~& ~~~~~ BR (Theo.) ~~~~~   &  BR (Exp.) & Cabibbo\\
\hline
$\tau^- \to \pi^{0}  \pi^{-} \nu_{\tau}  $ & ~~$2.48\times 10^{-2}$~~  &~~$(3.0 \pm 3.2) \times 10^{-3}$~~& ~~{\footnotesize allowed}  \\
\hline
$\tau^- \to \eta \pi^{-} \nu_{\tau}  $ & ~~$0$~~  & ~~$ < 9.9 \times 10^{-5}$ ~~& ~~{\footnotesize  allowed}  \\
\hline
$\tau^- \to \eta^{\prime} \pi^{-} \nu_{\tau}  $ & ~~$0$~~  & ~~$ < 4.0 \times 10^{-6}$ ~~& ~~{\footnotesize  allowed}  \\
\hline
$\tau^- \to \eta K^{-} \nu_{\tau}  $ & ~~$8.17\times 10^{-5}$~~ & ~~$(1.55 \pm 0.08) \times 10^{-4}$ ~~& ~~{\footnotesize suppressed}  \\
\hline
$\tau^- \to \eta^{\prime} K^{-} \nu_{\tau}  $ & ~~$  3.26  \times 10^{-7}$~~  &  ~~$ < 2.4 \times 10^{-6}$ ~~& ~~{\footnotesize suppressed} \\
\hline
$\tau^- \to \pi^{0} K^{-} \nu_{\tau}  $ & ~~$ 1.29   \times 10^{-4}$~~ & ~~$(2.7 \pm 1.1) \times 10^{-3}$~~& ~~{\footnotesize  suppressed}\footnote{The PDG has only the whole contribution including $K^{*-}$ production. We evaluate the rates in two ways: $\frac{1}{2}$ of the rate of $\tau^- \to \pi^{-}  \bar{K}^{0} \nu_{\tau}$ (non resonance)
and taking the whole range times the ratio of $BR(\tau^- \to \pi^{-}  \bar{K}^{0} \nu_{\tau} ~{\rm non~ resonance})/BR(\tau^- \to \pi^{-}  \bar{K}^{0} \nu_{\tau}~{\rm whole})$.
Both ways give the same result.
The error is taken from  $\tau^- \to \pi^{-}  \bar{K}^{0} \nu_{\tau}$ in the table. }\\
\hline
$\tau^- \to K^{-} K^{0} \nu_{\tau}  $ & ~ {\color{red}(fit to the Exp.)}~~  & ~~$(1.48 \pm 0.05) \times 10^{-3}$ ~~& ~~{\footnotesize allowed} \\
\hline
$\tau^- \to \pi^{-}  \bar{K}^{0} \nu_{\tau}  $ & ~~$2.52   \times 10^{-4}$~~  & ~~$(5.4 \pm 2.1) \times 10^{-4}$ ~~& ~~{\footnotesize suppressed} \\
 \midrule[1.0pt]\midrule[1.0pt]
 $\tau^- \to \pi^{-} \rho^{0} \nu_{\tau}  $ & ~ $3.90 \times 10^{-3}$~~  & ~~ ~~& ~~{\footnotesize  allowed} \\
\hline
$\tau^- \to \pi^{-}\omega \nu_{\tau}  $ & ~ $5.31 \times 10^{-3}$~~  & ~~$(1.95 \pm 0.06)\% $ ~~& ~~{\footnotesize  allowed} \\
\hline
 $\tau^- \to \pi^{0} \rho^{-} \nu_{\tau}  $ & ~ $3.95 \times 10^{-3}$~~  & ~~ ~~& ~~{\footnotesize  allowed} \\
\hline
 $\tau^- \to \eta \rho^{-} \nu_{\tau}  $ & ~$4.32 \times 10^{-4}$~~  & ~~ ~~& ~~{\footnotesize allowed} \\
\hline
 $\tau^- \to \eta^{\prime} \rho^{-} \nu_{\tau}  $ & ~ $8.25 \times 10^{-9}$~~  & ~~ ~~& ~~{\footnotesize  allowed} \\
\hline
 $\tau^- \to K^{0} K^{*-} \nu_{\tau}  $ & ~ $2.51 \times 10^{-4}$~~  & ~~ ~~& ~~{\footnotesize  allowed} \\
\hline
 $\tau^- \to K^{*0} K^{-}  \nu_{\tau}  $ & ~ $2.49  \times 10^{-4}$~~  & ~~$(2.1 \pm 0.4) \times 10^{-3}$ ~~& ~~{\footnotesize  allowed} \\
\hline
 $\tau^- \to  K^{-} \rho^{0} \nu_{\tau}  $ & ~ $2.18 \times 10^{-5}$~~  & ~~$(1.4 \pm 0.5) \times 10^{-3}$~~& ~~{\footnotesize  suppressed} \\
\hline
 $\tau^- \to  K^{-} \omega \nu_{\tau}  $ & ~ $2.04  \times 10^{-5}$~~  & ~~$(4.1 \pm 0.9) \times 10^{-4}$ ~~& ~~{\footnotesize suppressed} \\
\hline
$\tau^- \to  \bar{K}^{0}\rho^{-}  \nu_{\tau}  $ & ~~$4.22 \times 10^{-5}$~~  & ~~$(2.2 \pm 0.5) \times 10^{-3}$ ~~& ~~{\footnotesize suppressed} \\
\hline
 $\tau^- \to \eta K^{*-} \nu_{\tau}  $ & ~ $3.70 \times 10^{-6}$~~  & ~~$(1.38\pm 0.15) \times 10^{-4}$ ~~& ~~{\footnotesize  suppressed} \\
\hline
 $\tau^- \to \eta^{\prime}  K^{*-}\nu_{\tau}  $ & ~ $0$~~  & ~~ ~~& ~~{\footnotesize  suppressed} \\
\hline
 $\tau^- \to \pi^{0}  K^{*-}\nu_{\tau}  $ & ~ $6.37 \times 10^{-5}$~~  & ~~ ~~& ~~{\footnotesize suppressed} \\
\hline
 $\tau^- \to \bar{K}^{*0} \pi^{-}  \nu_{\tau}  $ & ~ $1.22 \times 10^{-4}$~~  & ~~$(2.2 \pm 0.5) \times 10^{-3}$ ~~& ~~{\footnotesize  suppressed} \\
\hline
 $\tau^- \to \phi K^{-}   \nu_{\tau}  $ & ~ $2.40 \times 10^{-6}$~~  & ~~$(4.4 \pm 1.6) \times 10^{-5}$ ~~& ~~{\footnotesize  suppressed} \\
\hline
 $\tau^- \to \rho^{-}  \rho^{0} \nu_{\tau}  $ & ~$1.24 \times 10^{-4}$~~  & ~~ ~~& ~~{\footnotesize suppressed} \\
\hline
 $\tau^- \to \rho^{-} \omega \nu_{\tau}  $ & ~$3.35 \times 10^{-5}$~~  & ~~ ~~& ~~{\footnotesize suppressed} \\
\hline
 $\tau^- \to K^{*0}  K^{*-}\nu_{\tau}  $ & ~ $0$~~  & ~~ ~~& ~~{\footnotesize suppressed} \\
\hline
 $\tau^- \to K^{*-} \rho^{0} \nu_{\tau}  $ & ~$9.04 \times 10^{-8}$~~  & ~~ ~~& ~~{\footnotesize  suppressed} \\
\hline
 $\tau^- \to K^{*-} \omega\nu_{\tau}  $ & ~$6.65 \times 10^{-8}$~~  & ~~ ~~& ~~{\footnotesize suppressed} \\
\hline
 $\tau^- \to \bar{K}^{*0} \rho^{-} \nu_{\tau}  $ & ~$1.54  \times 10^{-7}$~~  & ~~ ~~& ~~{\footnotesize  suppressed} \\
\hline
 $\tau^- \to K^{*-} \phi\nu_{\tau}  $ & ~ $0$~~  & ~~ ~~& ~~{\footnotesize  suppressed} \\
\hline
\bottomrule[1.0pt]\bottomrule[1.0pt]
\end{tabular}
\label{tab:p1}
\end{table}

\begin{table}[h!]
\renewcommand\arraystretch{0.92}
\caption{\color{black} Branching ratios for $\tau^- \to \nu_{\tau} M_1 M_2  $  in $s$-wave normalized by  $\tau^- \to \eta K^{*-} \nu_{\tau}$  }
\centering
\begin{tabular}{l  c  c  }
\toprule[1.0pt]\toprule[1.0pt]
{Decay process~} ~&~~~~~~~~ BR (Theo.) ~~~~~~~ &  BR (Exp.) \\
\hline
 $\tau^- \to \pi^{-} \rho^{0} \nu_{\tau}  $ & ~ $ 7.68 \times 10^{-2}$~~  & ~~ ~~ \\
\hline
$\tau^- \to \pi^{-}\omega \nu_{\tau}  $ & ~ $  5.80 \times 10^{-2}$~~  & ~~$(1.95 \pm 0.06)\% $ ~~ \\
\hline
 $\tau^- \to \pi^{0} \rho^{-} \nu_{\tau}  $ & ~ $ 7.78 \times 10^{-2}$~~  & ~~ ~~\\
\hline
 $\tau^- \to \eta \rho^{-} \nu_{\tau}  $ & ~$  4.50 \times 10^{-3}$~~  & ~~ ~~ \\
\hline
 $\tau^- \to \eta^{\prime} \rho^{-} \nu_{\tau}  $ & ~ $ 5.89 \times 10^{-7}$~~  & ~~ ~~ \\
\hline
 $\tau^- \to K^{0} K^{*-} \nu_{\tau}  $ & ~ $ 4.95 \times 10^{-3}$~~  & ~~ ~~ \\
\hline
 $\tau^- \to K^{*0} K^{-}  \nu_{\tau}  $ & ~ $ 4.93 \times 10^{-3}$ ~~  & ~~$(2.1 \pm 0.3) \times 10^{-3}$ ~~\\
\hline
 $\tau^- \to  K^{-} \rho^{0} \nu_{\tau}  $ & ~ $ 3.41 \times 10^{-4}$~~  & ~~$(1.4 \pm 0.5) \times 10^{-3}$~~ \\
\hline
 $\tau^- \to  K^{-} \omega \nu_{\tau}  $ & ~ $  3.24  \times 10^{-4}$~~  & ~~$(4.1 \pm 0.9) \times 10^{-4}$ ~~\\
\hline
$\tau^- \to  \bar{K}^{0}\rho^{-}  \nu_{\tau}  $ & ~~$ 6.64 \times 10^{-4}$~~  & ~~$(2.2 \pm 0.5) \times 10^{-3}$ ~~\\
\hline
 $\tau^- \to \eta K^{*-} \nu_{\tau}  $ & ~ {\color{red}(fit to the exp)}~~  & ~~$(1.38\pm 0.15) \times 10^{-4}$ ~\\
\hline
 $\tau^- \to \eta^{\prime}  K^{*-}\nu_{\tau}  $ & ~ $ 0$~~  & ~~ ~ \\
\hline
 $\tau^- \to \pi^{0}  K^{*-}\nu_{\tau}  $ & ~ $ 1.07 \times 10^{-3}$~~  & ~~ ~\\
\hline
 $\tau^- \to \bar{K}^{*0} \pi^{-}  \nu_{\tau}  $ & ~ $ 2.05 \times 10^{-3}$~~  & ~~$(2.2 \pm 0.5) \times 10^{-3}$ ~ \\
\hline
 $\tau^- \to \phi K^{-}   \nu_{\tau}  $ & ~ $ 6.82 \times 10^{-5}$~~  & ~~$(4.4 \pm 1.6) \times 10^{-5}$ ~~ \\
\hline\hline
 $\tau^- \to \rho^{-}  \rho^{0} \nu_{\tau}  $ & ~$ 1.15\times 10^{-3}$~~  & ~~ ~~ \\
\hline
 $\tau^- \to \rho^{-} \omega \nu_{\tau}  $ & ~$ 3.19 \times 10^{-3}$~~  & ~~ ~ \\
\hline
 $\tau^- \to K^{*0}  K^{*-}\nu_{\tau}  $ & ~ $ 0$~~~  & ~~ ~ \\
\hline
 $\tau^- \to K^{*-} \rho^{0} \nu_{\tau}  $ & ~$ 5.15 \times 10^{-6}$~~  & ~~  \\
\hline
 $\tau^- \to K^{*-} \omega\nu_{\tau}  $ & ~$  4.05 \times 10^{-6}$~~  & ~~ ~ \\
\hline
 $\tau^- \to \bar{K}^{*0} \rho^{-} \nu_{\tau}  $ & ~$ 9.09  \times 10^{-6}$~~  & ~~ \\
\hline
 $\tau^- \to K^{*-} \phi\nu_{\tau}  $ & ~ $0$~~  & ~~ ~ \\
\hline
\bottomrule[1.0pt]\bottomrule[1.0pt]
\end{tabular}
\label{tab:s1}
\end{table}

\begin{table}[h!]
\renewcommand\arraystretch{0.92}
\caption{\color{black} Branching ratios for $\tau^- \to \nu_{\tau} M_1 M_2$    in $s$-wave {\color{black} after convolution}  normalized by  $\tau^- \to \eta K^{*-} \nu_{\tau}$  }
\centering
\begin{tabular}{l  c  c  }
\toprule[1.0pt]\toprule[1.0pt]
{Decay process~} ~& ~~~~~~~~ BR (Theo.) ~~~~~~~~  &  BR (Exp.) \\
\hline
 $\tau^- \to \pi^{-} \rho^{0} \nu_{\tau}  $ & ~ $ 7.81 \times 10^{-2}$~~  & ~~ ~~ \\
\hline
$\tau^- \to \pi^{-}\omega \nu_{\tau}  $ & ~ $ 5.56 \times 10^{-2}$~~  & ~~$(1.95 \pm 0.06)\% $ ~~ \\
\hline
 $\tau^- \to \pi^{0} \rho^{-} \nu_{\tau}  $ & ~ $ 7.91 \times 10^{-2}$~~  & ~~ ~~\\
\hline
 $\tau^- \to \eta \rho^{-} \nu_{\tau}  $ & ~$ 5.34 \times 10^{-3}$~~  & ~~ ~~ \\
\hline
 $\tau^- \to \eta^{\prime} \rho^{-} \nu_{\tau}  $ & ~ $ 2.96 \times 10^{-5}$~~  & ~~ ~~ \\
\hline
 $\tau^- \to K^{0} K^{*-} \nu_{\tau}  $ & ~ $ 4.91 \times 10^{-3}$~~  & ~~ ~~ \\
\hline
 $\tau^- \to K^{*0} K^{-}  \nu_{\tau}  $ & ~ $ 4.87  \times 10^{-3}$~  & ~~$(2.1 \pm 0.3) \times 10^{-3}$ ~~\\
\hline
 $\tau^- \to  K^{-} \rho^{0} \nu_{\tau}  $ & ~ $ 3.82 \times 10^{-4}$~~  & ~~$(1.4 \pm 0.5) \times 10^{-3}$~~ \\
\hline
 $\tau^- \to  K^{-} \omega \nu_{\tau}  $ & ~ $  3.10  \times 10^{-4}$~~  & ~~$(4.1 \pm 0.9) \times 10^{-4}$ ~~\\
\hline
$\tau^- \to  \bar{K}^{0}\rho^{-}  \nu_{\tau}  $ & ~~$ 7.44 \times 10^{-4}$~~  & ~~$(2.2 \pm 0.5) \times 10^{-3}$ ~~\\
\hline
 $\tau^- \to \eta K^{*-} \nu_{\tau}  $ & ~  {\color{red}(fit to the Exp.)}~~  & ~~$(1.38\pm 0.15) \times 10^{-4}$ ~\\
\hline
 $\tau^- \to \eta^{\prime}  K^{*-}\nu_{\tau}  $ & ~ $ 1.21 \times 10^{-10}$~~  & ~~ ~ \\
\hline
 $\tau^- \to \pi^{0}  K^{*-}\nu_{\tau}  $ & ~ $ 1.03 \times 10^{-3}$~~  & ~~ ~\\
\hline
 $\tau^- \to \bar{K}^{*0} \pi^{-}  \nu_{\tau}  $ & ~ $ 1.99 \times 10^{-3}$~~  & ~~$(2.2 \pm 0.5) \times 10^{-3}$ ~ \\
\hline
 $\tau^- \to \phi K^{-}   \nu_{\tau}  $ & ~ $ 6.54 \times 10^{-5}$~~  & ~~$(4.4 \pm 1.6) \times 10^{-5}$ ~~ \\
\hline\hline
 $\tau^- \to \rho^{-}  \rho^{0} \nu_{\tau}  $ & ~$ 3.31 \times 10^{-3}$~~  & ~~ ~~ \\
\hline
 $\tau^- \to \rho^{-} \omega \nu_{\tau}  $ & ~$  5.82 \times 10^{-3}$~~  & ~~ ~ \\
\hline
 $\tau^- \to K^{*0}  K^{*-}\nu_{\tau}  $ & ~ $ 8.18 \times 10^{-6}$~~~  & ~~ ~ \\
\hline
 $\tau^- \to K^{*-} \rho^{0} \nu_{\tau}  $ & ~$ 2.96 \times 10^{-5}$~~  & ~~  \\
\hline
 $\tau^- \to K^{*-} \omega\nu_{\tau}  $ & ~$  6.0  \times 10^{-6}$~~  & ~~ ~ \\
\hline
 $\tau^- \to \bar{K}^{*0} \rho^{-} \nu_{\tau}  $ & ~$ 5.46  \times 10^{-5}$~~  & ~~ \\
\hline
 $\tau^- \to K^{*-} \phi\nu_{\tau}  $ & ~ $0$~~  & ~~ ~ \\
\hline
\bottomrule[1.0pt]\bottomrule[1.0pt]
\end{tabular}
\label{tab:s2}
\end{table}

\section{Conclusions}
\label{sec:conc}
We have performed a study of the $\tau^-$ decay into $\nu_{\tau}$ and two mesons, with the aim of establishing a relationship between  production of two
pseudoscalars, a pseudoscalar and a vector and two vectors. For this we have used the dynamics  of the weak interaction and worked out  all the
angular monmentum-spin algebra to relate these processes, provided  the form factors stemming from the radial  wave functions are the same in the different cases, which involve only
quarks in their ground states.

 The calculations done allow us to present a new perspective  of the role played by $G$-parity in these reactions, involoving $u,d$ quarks. However, we also find
  that the selection rules  of   $G$-parity have repercussion in the matrix elements of $\tau^- \to K^{-} \eta \nu_{\tau}$, $\tau^- \to K^{-} \eta' \nu_{\tau}$,
 $\tau^- \to K^{*-} \eta \nu_{\tau}$, $\tau^- \to K^{*-} \eta' \nu_{\tau}$, where  $G$-parity does not apply.

 We compare  our results with experiment. For  $\tau^- $ decays into $\nu_{\tau}$ and two pseudoscalars we assume  that the two mesons are produced with $p$-wave.
 This is agreement with the formalism of chiral perturbation theory. In our case the two mesons are produced from an initial $q\bar{q}$ formation  by the $W$, followed
 by the hadronization  of $q\bar{q}$ into two mesons, which is  done using the $^3P_0$ model.

 However, we observe that assuming also $p$-wave for the pseudoscalar-vector and vector-vector  production one obtains results clearly incompatible with experimental data.
This fact and experimental evidence that in such case the mesons are produced  in $s$-wave, leads us to redo the formalism  for production of the two mesons in $s$-wave.

 Comparison with the experimental results shows that our predicitons are fair, in spite of the large differences in the rates  for different  cases. We also make predictions
 for unmeasured  decays.

 Another point in the results is that sometimes there are larger discrepancies from the data, and in these cases we could identify the reason of the discrepancies
 to large resonance contribution, with the resonance decaying finally into the two meson observed.

  We also emphasize that our formalism can be directly used to  take into account final state interaction of the mesons that in some cases lead to dynamically
  generated resonances.

  Finally we also emphasize the value of these decays to study the meson-meson interaction  and the nature of some resonances, which should stimulate experimentalists
  to measure the two-meson mass distributions in these decays in  analogy to  what is done in the  $\tau^- \to \nu_{\tau} a_1(1260) \to \nu_{\tau} \pi \rho$,
  where the $\pi \rho$ mass distribution is measured with great precision.

\section*{Acknowledgements}
L. R. Dai wishes to acknowledge the support from the State Scholarship Fund of China (No. 201708210057)
and the National Natural Science Foundation of China (No. 11575076).  R. P. Pavao wishes to
thank the Generalitat Valenciana in the program Santiago Grisolia.
This work is partly supported by the Spanish Ministerio de Economia y Competitividad
and European FEDER funds under the contract number FIS2011-
28853-C02-01, FIS2011- 28853-C02-02, FIS2014-57026-
REDT, FIS2014-51948-C2- 1-P, and FIS2014-51948-C2-
2-P, and the Generalitat Valenciana in the program
Prometeo II-2014/068 (EO).

\appendix
\section{Evaluation of the matrix elements for the operators $"1"$ and $\sigma_i$}
We start from Eq.~\eqref{eq:ME}
\begin{equation*}
\text{ME} = -\frac{1}{\sqrt{3}} \sum_{S_3} \sum_{s} (-1)^{\frac{1}{2}-s-M'} \mathcal{C}\left(\frac{1}{2} \frac{1}{2} 1; s, S_3-s,S_3\right)  \ q  \ Y_{1,-S_3} ({\hat{\bm{q}}}) \ \mathcal{C}\left(\frac{1}{2} \frac{1}{2} J, M-s,s,M\right)
\end{equation*}
\begin{equation}
\times \mathcal{C}\left(\frac{1}{2} \frac{1}{2} J', S_3-s,M'-S_3+s,M'\right) \cdot \left\{\begin{matrix}
\! \! \! \!  \left < m  \right | 1 \left | m' \right > \\
\left < m  \right | \sigma_{\mu} \left | m' \right >
\end{matrix}\right. ,
\end{equation}
where in the case of the $"1"$ operator we have $m=m'=M-s$ and $S_3=M+M'$, while in the case of $\sigma_{\mu}$ we have $m=M-s$, $m'+\mu=m$, $\mu = M-S_3+M'$. In the case of the operator $"1"$ we obtain $M_0$ and with $\sigma_{\mu}$, $N_{\mu}$.

\begin{enumerate}
\item[1)] $\mathbf{M_0}$:\\
We have  \begin{equation*}
-\sqrt{\frac{1}{3}} \sum_{s} (-1)^{\frac{1}{2}-s-M'}  \ q  \ Y_{1,-\left(M+M'\right)} ({\hat{\bm{q}}}) \ \mathcal{C}\left( \frac{1}{2} \frac{1}{2} 1; s, M+M'-s, M+M' \right)
\end{equation*}
\begin{equation}
\times  \mathcal{C}\left( \frac{1}{2} \frac{1}{2} J; M-s,s,M \right)  \mathcal{C}\left( \frac{1}{2} \frac{1}{2} J'; M+M'-s,-M+s,M' \right).
\end{equation}
Using the permutation relations \cite{rose}
\begin{subequations}
\begin{align}
& \mathcal{C}\left( \frac{1}{2} \frac{1}{2} 1; s, M+M'-s , M+M'\right)  = (-1)^{\frac{1}{2}-s} \sqrt{\frac{3}{2}} \ \mathcal{C}\left( 1 \frac{1}{2} \frac{1}{2} ; M+M',-s,M+M'-s \right), \\
& \mathcal{C}\left( \frac{1}{2} \frac{1}{2} J; M-s,s,M \right) = \mathcal{C}\left( \frac{1}{2} \frac{1}{2} J; -s,-M+s,-M \right),
\end{align}
\end{subequations}
we obtain:
\begin{equation*}
M_0 = -\frac{1}{\sqrt{3}} (-1)^{-M'} \sqrt{\frac{3}{2}} \ q  \ Y_{1,-\left(M+M'\right)}({\hat{\bm{q}}}) \sum_s  \mathcal{C}\left( 1 \frac{1}{2} \frac{1}{2} ; M+M',-s,M+M'-s \right)
\end{equation*}
\begin{equation}
\times \mathcal{C}\left( \frac{1}{2} \frac{1}{2} J'; M+M'-s,-M+s,M' \right) \mathcal{C}\left( \frac{1}{2} \frac{1}{2} J; -s,-M+s,-M \right),
\end{equation}
and summing over $s$, keeping $M$ fixed we obtain, using the formulas of \cite{rose},
\begin{equation}
M_0= -\frac{1}{\sqrt{2}} (-1)^{-M'} \sqrt{2(2J+1)} \ q  \ Y_{1,-\left(M+M'\right)}({\hat{\bm{q}}})  \ \mathcal{W}\left(1 \frac{1}{2} J' \frac{1}{2}; \frac{1}{2} J \right) \mathcal{C}\left( 1 J J'; M+M',-M,M' \right),
\end{equation}
in terms of a Racah coefficient, $\mathcal{W}(\cdots)$.

We can write this in a more symmetrical way by taking
\begin{equation}
\mathcal{C}\left( 1 J J'; M+M',-M,M' \right) = (-1)^{J-M} \sqrt{\frac{2J'+1}{3}} \ \mathcal{C}\left( J J' 1; M,M',M+M' \right),
\end{equation}
such that finally we obtain,
\begin{equation*}
M_0 = -(-1)^{J-(M+M')} \frac{1}{\sqrt{3}} \sqrt{(2J+1)(2J'+1)} \ \mathcal{W}\left(1 \frac{1}{2} J' \frac{1}{2}; \frac{1}{2} J \right)
\end{equation*}
\begin{equation}
\times \ \mathcal{C}\left( J J' 1; M,M',M+M' \right)  \ q \ Y_{1,-\left(M+M'\right)} ({\hat{\bm{q}}}).
\end{equation}
We apply it to the different $M_1 M_2$ cases:
\begin{itemize}
\item[a)] $PP: \ J=0, J'=0$

The Clebsch-Gordan Coefficient (CGC) $\mathcal{C}\left( 0 0 1; \cdots \right)$ is zero, hence:
\begin{equation}
M_0= 0.
\end{equation}

\item[b)] $PV: \ J=0, J'=1$

Using the table in the Appendix of \cite{rose} we find
\begin{equation}
\mathcal{W}\left(1 \frac{1}{2} 1 \frac{1}{2}; \frac{1}{2} 0 \right) = -\frac{1}{\sqrt{6}},
\end{equation}
and then, for any $M'$,
\begin{equation}
M_0 = (-1)^{-M-M'} \frac{1}{\sqrt{6}} \delta_{M 0} \ q \ Y_{1,-\left(M+M'\right)} ({\hat{\bm{q}}})
\end{equation}

\item[c)] $VP: \ J=1, J'=0$

Now
\begin{equation}
\label{eq:3-1}
\mathcal{W}\left(1 \frac{1}{2} 0 \frac{1}{2}; \frac{1}{2} 1 \right) = \frac{1}{\sqrt{6}},
\end{equation}
and thus we get, for any $M$,
\begin{equation}
M_0 = (-1)^{-M-M'} \frac{1}{\sqrt{6}} \delta_{M' 0} \ q \ Y_{1,-\left(M+M'\right)} ({\hat{\bm{q}}})
\end{equation}

\item[d)] $VV: \ J=1, J'=1$

Now
\begin{equation}
\label{eq:4-1}
\mathcal{W}\left(1 \frac{1}{2} 1 \frac{1}{2}; \frac{1}{2} 1 \right) = \frac{1}{3},
\end{equation}
and thus we get
\begin{equation}\\
\label{eq:4-2}
M_0 = (-1)^{-M-M'} \frac{1}{\sqrt{3}} \mathcal{C}\left(1 1 1; M,M',M+M'\right) \ q \ Y_{1,-\left(M+M'\right)} ({\hat{\bm{q}}}).
\end{equation}
\end{itemize}

\item[2)] $\mathbf{N_{\mu}}$:\\

We have:
\begin{equation*}
N_{\mu} = -\frac{1}{\sqrt{3}} \sum_s (-1)^{\frac{1}{2}-s-M'} \ q \ Y_{1,\mu-(M+M')} ({\hat{\bm{q}}}) \ \mathcal{C}\left(\frac{1}{2} \frac{1}{2} 1; s, M+M'-\mu -s, M+M'- \mu \right)
\end{equation*}
\begin{equation*}
\times \mathcal{C}\left( \frac{1}{2} \frac{1}{2} J; M-s,s,M\right) \mathcal{C}\left( \frac{1}{2} \frac{1}{2} J'; M+M'-\mu -s,-M+\mu+s,M'\right)
\end{equation*}
\begin{equation}
\times \sqrt{3} \  \mathcal{C}\left( \frac{1}{2} 1 \frac{1}{2}; M-s-\mu,\mu, M-s\right).
\end{equation}

Note that now the variable $s$ is in the four CGC and we cannot get directly a Racah coefficient. For this we use again formulas of \cite{rose} to decompose two CGC into other two, one of which does not depend on s. First we use the permutations
\begin{equation*}
\mathcal{C}\left(\frac{1}{2} \frac{1}{2} 1; s, M+M'-\mu -s, M+M'- \mu \right)
\end{equation*}
\begin{equation}
= (-1)^{\frac{1}{2}-s} \sqrt{\frac{3}{2}}  \ \mathcal{C}\left( 1 \frac{1}{2} \frac{1}{2}; M+M'- \mu,-s, M+M'-\mu -s \right),
\end{equation}
and
\begin{equation}
\mathcal{C}\left( \frac{1}{2} \frac{1}{2} J; M-s,s,M\right) = \mathcal{C}\left( \frac{1}{2} \frac{1}{2} J; -s,-M+s,-M\right),
\end{equation}
and we find:
\begin{equation*}
N_{\mu} = -\frac{1}{\sqrt{3}} \sum_s (-1)^{-M'} \ q \ Y_{1,\mu-(M+M')} ({\hat{\bm{q}}}) \  \sqrt{\frac{3}{2}} \ \mathcal{C}\left( 1 \frac{1}{2} \frac{1}{2}; M+M'- \mu,-s, M+M'-\mu -s \right)
\end{equation*}
\begin{equation*}
\times \mathcal{C}\left( \frac{1}{2} \frac{1}{2} J'; M+M'-\mu -s,-M+\mu+s,M'\right) \mathcal{C}\left( \frac{1}{2} \frac{1}{2} J; -s,-M+s,-M\right)
\end{equation*}
\begin{equation}
\times  \sqrt{3} \  \mathcal{C}\left( \frac{1}{2} 1 \frac{1}{2}; M-s-\mu,\mu, M-s\right).
\end{equation}
We can use formulas of \cite{rose} and write the first two CGC as
\begin{equation*}
\mathcal{C}\left( 1 \frac{1}{2} \frac{1}{2}; M+M'- \mu,-s, M+M'-\mu -s \right) \mathcal{C}\left( \frac{1}{2} \frac{1}{2} J'; M+M'-\mu -s,-M+\mu+s,M'\right)
\end{equation*}
\begin{equation*}
 = \sum_{j''} \sqrt{2(2j''+1)}  \mathcal{W}\left(1 \frac{1}{2} J' \frac{1}{2}; \frac{1}{2} j'' \right)\mathcal{C}\left( \frac{1}{2} \frac{1}{2}  j''; -s, -M+\mu+s,-M+\mu \right)
\end{equation*}
\begin{equation}
\times \mathcal{C}\left( 1 \ j'' J'; M+M'-\mu , -M+\mu , M' \right).
\end{equation}
We use again CGC permutation relations:
\begin{subequations}
\label{eq:5-1}
\begin{align}
&  \mathcal{C}\left( \frac{1}{2} 1 \frac{1}{2}; M-s-\mu,\mu, M-s\right) = -  \mathcal{C}\left( 1 \frac{1}{2} \frac{1}{2}; \mu,M-s-\mu, M-s\right), \\
& \mathcal{C}\left( \frac{1}{2} \frac{1}{2} J; -s,-M+s,-M\right) = \mathcal{C}\left( \frac{1}{2} \frac{1}{2} J; M-s, s,M\right), \\
& \mathcal{C}\left( \frac{1}{2} \frac{1}{2}  j''; -s, -M+\mu+s,-M+\mu \right) = \mathcal{C}\left( \frac{1}{2} \frac{1}{2}  j''; M-\mu-s,s, M-\mu \right).
\end{align}
\end{subequations}
Then summing over $M-\mu-s$, keeping $M-\mu$ fixed, we get for the sum of the three CGC to the right of Eq.~\eqref{eq:5-1} \cite{rose}
\begin{equation}
\sqrt{2(2j''+1)} \  \mathcal{W}\left(1 \frac{1}{2} J \frac{1}{2}; \frac{1}{2} j'' \right) \mathcal{C}\left( 1  j'' J; \mu,M-\mu, M \right).
\end{equation}
So, finally we get
\begin{equation*}
N_{\mu} = (-1)^{-M'} \sqrt{6} \ q \ Y_{1,\mu-(M+M')} (\mathbf{\hat{q}}) \sum_{j''} (2j''+1)  \  \mathcal{W}\left(1 \frac{1}{2} J \frac{1}{2}; \frac{1}{2} j'' \right) \mathcal{W}\left(1 \frac{1}{2} J' \frac{1}{2}; \frac{1}{2} j'' \right)
\end{equation*}
\begin{equation}
\times \mathcal{C}\left( 1  j'' J; \mu,M-\mu, M \right) \mathcal{C}\left( 1 \ j'' J'; M+M'-\mu , -M+\mu , M' \right).
\end{equation}
We apply this equation to the different $M_1 M_2$ cases and find:

\begin{itemize}
\item[a)] $PP: \ J=0, J'=0$

\begin{equation}
\label{eq:6-1}
\mathcal{C}\left( 1 j'' 0; \mu, M-\mu, M \right) = (-1)^{1-\mu} \sqrt{\frac{1}{2j''+1}} \ \mathcal{C}\left( 1 0 j''; \mu, -M, \mu-M \right),
\end{equation}
which implies $M=0$, and $j''=1$,
\begin{equation*}
\mathcal{C}\left( 1 j'' 0; M+M'-\mu, -M+\mu,M' \right) = (-1)^{1-M-M'+\mu} \sqrt{\frac{1}{2j''+1}}
\end{equation*}
\begin{equation}
\times \mathcal{C}\left( 1 0 j'';  M+M'-\mu, -M',M-\mu \right),
\end{equation}
which also implies that $M'=0$ and $j''=1$. The Racah coefficients are the same as in Eq.~\eqref{eq:3-1} and we finally get
\begin{equation}
\label{eq:6-2}
N_{\mu} = \frac{1}{\sqrt{6}} \ q \ Y_{1,\mu} ({\hat{\bm{q}}}) \ \delta_{M0} \delta_{M'0}.
\end{equation}

\item[b)] $PV: \ J=0, J'=1$

We use Eq.~\eqref{eq:6-1} which implies $M=0$ and $j''=1$ and then write
\begin{equation}
\mathcal{C}\left( 1 1 1; M+M'-\mu, -M+\mu,M' \right) =(-1)^{1-M-M'+\mu} \ \mathcal{C}\left( 1 1 1; M+M'-\mu, -M', M-\mu \right).
\end{equation}
We need the Racah coefficients of Eqs.~\eqref{eq:3-1} and \eqref{eq:4-1}, and we get
\begin{equation}
N_{\mu} = (-1)^{-M} \frac{1}{\sqrt{3}} \ q \ Y_{1,\mu-(M+M')} ({\hat{\bm{q}}}) \  \delta_{M 0} \
 \mathcal{C}\left( 1 1 1; M'-\mu , -M', -\mu  \right),
\end{equation}
and writing the CGC as $(-1)^{1-M'} \mathcal{C}\left( 1 1 1; M',-\mu , M'- \mu  \right)$ we get finally
\begin{equation}
\label{eq:A28}
N_{\mu} = (-1)^{1-M-M'} \frac{1}{\sqrt{3}} \ q \ Y_{1,\mu-M'} ({\hat{\bm{q}}}) \  \delta_{M 0} \ \mathcal{C}\left( 1 1 1; M', -\mu , M'- \mu  \right).
\end{equation}

\item[c)] $VP: \ J=1, J'=0$

We use
\begin{equation*}
\mathcal{C}\left( 1 j'' 0; M+M'-\mu, -M+\mu, M' \right) = (-1)^{1-M-M'+\mu} \sqrt{\frac{1}{2j''+1}}
\end{equation*}
\begin{equation}
\times \mathcal{C}\left( 1 0 j''; M+M'-\mu, -M', M-\mu \right),
\end{equation}
which implies $M'=0$ and $j''=1$ and using
\begin{equation}
\mathcal{C}\left( 1 1 1; \mu, -M, \mu-M \right) = \mathcal{C}\left( 1 1 1; M, -\mu, M-\mu \right),
\end{equation}
we finally find
\begin{equation}
N_{\mu} = (-1)^{-M} \frac{1}{\sqrt{3}} \ q \ Y_{1,\mu-M} ({\hat{\bm{q}}}) \  \delta_{M' 0} \ \mathcal{C}\left( 1 1 1; M, -\mu, M-\mu \right).
\end{equation}

\item[d)] $VV: \ J=1, J'=1$

We find now \cite{rose}
\begin{equation}
\mathcal{W}^2 \left(1 \frac{1}{2} 1 \frac{1}{2}; \frac{1}{2} j'' \right) = \frac{1}{36} (3+j'') (2-j''),
\end{equation}
which means that only $j''=0$, $j''=1$ contribute and for $j''=2$ the coefficient is zero.

\begin{itemize}
\item[i)] If $j''=0$ we get
\begin{equation}
N_{\mu} \rightarrow (-1)^{-M'} \frac{1}{\sqrt{6}} \ \delta_{M \mu} \ q \ Y_{1,\mu-(M+M')} ({\hat{\bm{q}}}).
\end{equation}

\item[ii)] If $j''=1$ we write

\begin{subequations}
\begin{align}
& \mathcal{C}\left( 1 1 1; \mu , M-\mu, M \right) = (-1)^{1-\mu}  \mathcal{C}\left( 1 1 1; M, -\mu, M-\mu \right),\\
&  \mathcal{C}\left( 1 1 1; M+M'-\mu, -M+\mu, M' \right) = (-1)^{1-M-M'+\mu} \ \mathcal{C}\left( 1 1 1; M', -M-M'+\mu, -M+\mu \right),
\end{align}
\end{subequations}
and then
\begin{equation*}
N_{\mu} \rightarrow (-1)^{-M} \sqrt{\frac{2}{3}} \ q \ Y_{1,\mu-(M+M')}({\hat{\bm{q}}})	\ \mathcal{C}\left( 1 1 1; M, -\mu, M-\mu \right)
\end{equation*}
\begin{equation}
\times \mathcal{C}\left( 1 1 1; M', -M-M'+\mu, -M+\mu \right),
\end{equation}
\end{itemize}
and for the sum of $j''=0$, $j''=1$  we get the final result
\begin{equation*}
N_{\mu}=  \frac{1}{\sqrt{6}}  \ q \ Y_{1,\mu-(M+M')} ({\hat{\bm{q}}}) \Big\{(-1)^{-M'} \ \delta_{M \mu} + 2 (-1)^{-M} \ \mathcal{C}\left( 1 1 1; M, -\mu, M-\mu \right)
\end{equation*}
\begin{equation}
\label{eq:8-1}
 \times \mathcal{C}\left( 1 1 1; M', -M-M'+\mu, -M+\mu \right) \Big\}
\end{equation}

\end{itemize}
\end{enumerate}

\section{Evaluation of $\overline{\sum} \sum |t|^2$}

Following the nomenclature $\bar{L}^{\mu \nu} = \overline{\sum} \sum L^{\mu} L^{\dagger \nu}$ adopted before for simplicity, we have for the leptonic sector
\begin{equation}
\bar{L}^{\mu \nu} \equiv  \overline{\sum} \sum L^{\mu} L^{\dagger \nu} = \frac{1}{m_{\tau} m_{\nu}} \left\{ p'^{\mu} p^{\nu} + p'^{\nu} p^{\mu} - g^{\mu \nu} (p' \cdot p) + i \epsilon^{\alpha \mu \beta \nu} p'_{\alpha} p_{\beta} \right\}.
\end{equation}
Thus for the leptonic plus hadronic matrix elements we have
\begin{equation}
\label{eq:9-1}
\overline{\sum} \sum |t|^2 = \bar{L}^{0 0} M_0 M^*_0 + \bar{L}^{0 i} M_0 N^*_i + \bar{L}^{i 0} N_i M^*_0 +  \bar{L}^{i j} N_i N^*_j.
\end{equation}
We have to take the product of these hadronic components, sum over $M,M'$ and contract with $\bar{L}^{\mu \nu}$. We do that for the different $M_1 M_2$ cases.

\begin{itemize}
\item[a)] $PP: \ J=0, J'=0$

In this case $M_0 = 0$ and we only have to calculate $N_i N^*_j$.

We use Eq.\eqref{eq:6-2} and write
\begin{equation}
q \ Y_{1,\mu}({\hat{\bm{q}}}) =  \sqrt{\frac{3}{4 \pi}} q_{\mu} = \sqrt{\frac{3}{4 \pi}} (\tilde{p}_1-\tilde{p}_2) = \sqrt{\frac{3}{4 \pi}} 2 \tilde{p}_1,
\end{equation}
since $\tilde{p}_1$ is evaluated in the rest frame of $M_1 M_2$. This means that in cartesian coordinates we can write
\begin{equation}
N_i= \frac{1}{\sqrt{6}} \sqrt{\frac{3}{4 \pi}} \ \delta_{M 0} \delta_{M' 0} \ 2 \tilde{p}_{1 i},
\end{equation}
and then from Eq.~\eqref{eq:9-1}:
\begin{equation}
\overline{\sum} \sum |t|^2 = \sum_{M M'} \frac{1}{m_{\tau} m_{\nu}} \frac{1}{2 \pi} \left \{2 p_i p_j + \delta_{ij} (p\cdot p') \right\} \tilde{p}_{1 i} \tilde{p}_{1 j}  \delta_{M 0} \delta_{M' 0},
\end{equation}
with
\begin{equation}
p_i p_j  \tilde{p}_{1 i} \tilde{p}_{1 j} =\left(\bm{p} \cdot {\tilde{\bm{p}}}_1 \right)^2= \left(p \tilde{p}_1 \right)^2 \cos^2 \theta \rightarrow \frac{1}{3} \left(p \tilde{p}_1 \right)^2,
\end{equation}
where the last step comes from the integral over $\cos^2 \theta$.
We replace $\cos^2 \theta$ by $1/3$ and put the whole phase space later independent on the angles. Then we get, including the weight $\bar{h}_i$ for the $N_i$ term
\begin{equation}
\label{eq:B7}
\overline{\sum} \sum |t|^2 = \frac{1}{m_{\tau} m_{\nu}} \frac{1}{2 \pi} \left( E_{\tau} E_{\nu} - \frac{1}{3} {\bm{p}}^2  \right) \tilde{p}_1^2 \bar{h}_i^2,
\end{equation}
with
\begin{equation}
\tilde{p}_1 = \frac{\lambda^{1/2} \left(M_{\text{inv}}^2(M_1M_2), M_1^2, M_2^2 \right)}{2 M_{\text{inv}}(M_1M_2)},
\end{equation}
and $p$ given in Eq.~\eqref{eq:newlabel},
\begin{equation}
p = \frac{\lambda^{1/2} \left(m_{\tau}^2, m_{\nu}^2, M_{\text{inv}}^2(M_1M_2) \right)}{2 M_{\text{inv}}(M_1M_2)}.
\end{equation}

In Eq.~\eqref{eq:B7} and what follows $E_{\tau}$, $E_{\nu}$ are also calculated in the $M_1 M_2$ rest frame, $E_{\tau} = \sqrt{m_{\tau}^2 + p^2}$, $E_{\nu}= p$.

The $\epsilon^{\alpha i \beta j} \tilde{p}_{1 i} \tilde{p}_{1 j}$ is zero. This term does not contribute in any case, but in some cases the cancellation comes from different terms in the sum over $M,M'$. The cancellation of this term when summing over polarizations in semileptonic decays was already found in \cite{Sekihara,Ikeno} and we do not elaborate on it further here.

\item[b)] $PV: \ J=0, J'=1$

Now we have
\begin{subequations}
\begin{align}
& M_0 = (-1)^{-M'} \frac{1}{\sqrt{6}} \delta_{M0} \ q \ Y_{1, -M'} ({\hat{\bm{q}}}),   \\
& N_{\mu} =  (-1)^{1-M'} \frac{1}{\sqrt{3}} \delta_{M0}  \ q \ Y_{1, \mu-M'} ({\hat{\bm{q}}}) \ \mathcal{C} \left(1 1 1; M', -\mu, M'-\mu \right).
\end{align}
\end{subequations}

First let us see that the $M_0 N_{\mu}$ components do not contribute. Indeed we find in the phase space integration
\begin{equation}
\sum_{M'} \int d \Omega Y_{1, -M'} Y^*_{1, \mu-M'} \ \mathcal{C} \left(1 1 1; M', -\mu, M'-\mu \right) = \delta_{\mu 0} \sum_{M'} \mathcal{C} \left(1 1 1; M', -\mu, M'-\mu \right) = 0.
\end{equation}
This is again the case also in $VP$ and $VV$ and we do not discuss it further.

Thus, we have contributions from:

\begin{itemize}
\item[i)] $M_0 M^*_0$

\begin{equation}
M_0 M^*_0 = \frac{1}{6} \delta_{M 0} Y_{1, -M'} ({\hat{\bm{q}}}) Y^*_{1, -M'} ({\hat{\bm{q}}}) \ q^2.
\end{equation}

In the phase space calculation we shall have
\begin{equation}
\int d\Omega  Y_{1, -M'} ({\hat{\bm{q}}}) Y^*_{1, -M'} ({\hat{\bm{q}}}) = 1,
\end{equation}
and then we replace $Y_1 Y_1^*$ by $\frac{1}{4 \pi}$ evaluating later the phase space for an angle independent amplitude.
Thus summing over $M' \ (M=0)$ we get
\begin{equation}
\sum_{MM'} M_0 M^*_0 = \frac{1}{6} \frac{1}{4 \pi} 4 \tilde{p}^2_1 \ 3 = \frac{1}{2 \pi} \tilde{p}_1^2,
\end{equation}
which multiplied by $\bar{L}^{00}$ gives
\begin{equation}
\label{eq:12-1}
\overline{\sum} \sum |t|^2 = \frac{1}{m_{\tau} m_{\nu}} \left(E_{\tau} E_{\nu} + {\bm{p}}^2  \right) \frac{1}{2 \pi} \tilde{p}_1^2.
\end{equation}

\item[ii)] $\bar{L}^{ij} N_i N^*_j$

For simplicity of the calculation we take $\mathbf{p}$ in the $z$ direction. Then
\begin{equation}
\bar{L}^{ij} N_i N^*_j =  \frac{1}{m_{\tau} m_{\nu}} \left( 2 p^2 \delta_{i 3} \delta_{j 3} N_i N^*_j + (p \cdot p') N_i N_i^* \right),
\end{equation}
but one can see that
\begin{equation}
\sum N_i N^*_i = \sum_{\mu} N_{\mu} N_{\mu}^*.
\end{equation}
Then, from Eq.~\eqref{eq:A28}
\begin{equation*}
\sum_{M,M'} \sum_{\mu} N_{\mu} N_{\mu}^* = \sum_{M'} \sum_{\mu} \frac{1}{4 \pi}  \frac{1}{3} q^2 \int Y_{1, \mu-M'} (\mathbf{\hat{q}}) Y^*_{1,\mu-M'} (\mathbf{\hat{q}}) d \Omega  \ \mathcal{C} \left(1 1 1; M', -\mu, M'-\mu \right)^2
\end{equation*}
\begin{equation*}
= \sum_{\mu, M'} \frac{1}{4 \pi} \frac{1}{3} 4 \tilde{p}_1^2 \ \mathcal{C} \left(1 1 1; M', -\mu, M'-\mu \right)^2 = \sum_{\mu, M'}  \frac{1}{3}  \frac{1}{4 \pi} \ \mathcal{C} \left(1 1 1; M',\mu-M',\mu \right)^2  4 \tilde{p}_1^2
\end{equation*}
 \begin{equation}
= \sum_{\mu} \frac{1}{4 \pi} \frac{4}{3} \tilde{p}_1^2= \frac{1}{ \pi} \tilde{p}_1^2.
 \end{equation}

On the other hand for $i=3$, $N_i \equiv N_{\mu = 0}$, and again
\begin{equation*}
\sum_{M'}  \frac{1}{4 \pi}  \frac{1}{3} q^2 \int Y_{1, -M'} ({\hat{\bm{q}}}) Y^*_{1,-M'} ({\hat{\bm{q}}}) d \Omega \ \mathcal{C} \left(1 1 1; M', 0 ,M' \right)^2
\end{equation*}
\begin{equation}
=\frac{1}{4 \pi} \frac{1}{3} 4 \tilde{p}_1^2 \sum_{M'} \mathcal{C} \left(1 1 1; M',-M',0 \right)^2 = \frac{1}{3 \pi}  \tilde{p}_1^2,
\end{equation}
and we find for this term
\begin{equation}
\label{eq:12-2}
\overline{\sum} \sum |t|^2  = \frac{1}{m_{\nu} m_{\tau}} \frac{1}{\pi} \tilde{p}_1^2 \left(E_{\tau} E_{\nu} - \frac{1}{3} {\bm{p}}^2  \right).
\end{equation}
\end{itemize}

Recalling that we have different weights for $M_0$ and $N_i$ in each channel we sum the two terms of Eqs.~\eqref{eq:12-1} and~\eqref{eq:12-2} to give
\begin{equation}
\overline{\sum} \sum |t|^2 = \frac{1}{m_{\tau} m_{\nu}} \frac{1}{2 \pi} \tilde{p}_1^2 \left\{h_i^2 \left(E_{\tau} E_{\nu} + {\bm{p}}^2  \right) + \bar{h}_i^2 \ 2   \left(E_{\tau} E_{\nu} - \frac{1}{3} {\bm{p}}^2  \right)  \right\}.
\end{equation}

\item[c)]  $VP: \ J=1, J'=0$

The evaluation proceeds as before and we obtain the same result.

\item[d)]  $VV: \ J=1, J'=1$

From Eqs.~\eqref{eq:4-2} and~\eqref{eq:8-1} we have
\begin{subequations}
\begin{align}
& M_0 = (-1)^{-M-M'} \frac{1}{\sqrt{3}} \  \mathcal{C} \left(1 1 1; M,M',M+M' \right) \ q \ Y_{1,-(M+M')} ({\hat{\bm{q}}}), \\
& \begin{matrix}
N_{\nu} = \frac{1}{\sqrt{6}} \ q \ Y_{1,\mu-(M+M')} ({\hat{\bm{q}}}) \left\{(-1)^{-M'} \delta_{M \mu} + 2  (-1)^{-M}  \mathcal{C} \left(1 1 1; M,-\mu,M-\mu \right) \right.\\
\times\left. \mathcal{C} \left(1 1 1; M',-M-M'+\mu,-M+\mu \right) \right\}.
\end{matrix}
\end{align}
\end{subequations}

\begin{itemize}

\item[i)] $M_0 M^*_0$

\begin{subequations}
\begin{align}
& \frac{1}{4 \pi} \int d \Omega  Y_{1,-(M+M')} ({\hat{\bm{q}}}) Y^*_{1,-(M+M')} ({\hat{\bm{q}}}) = \frac{1}{4 \pi}, \\
& \sum_{M,M'} \mathcal{C} \left(1 1 1; M,M',M+M' \right)^2 = \sum_{M,M'} \mathcal{C} \left(1 1 1; M,-M'-M,-M' \right)^2 = \sum_{M'} 1 = 3.
\end{align}
\end{subequations}
Then we get for this term
\begin{equation}
\overline{\sum} \sum |t|^2 = \frac{1}{m_{\tau} m_{\nu}} \frac{1}{\pi} \left(E_{\tau} E_{\nu} + {\bm{p}}^2  \right) \tilde{p}_1^2.
\end{equation}

\item[ii)] $\bar{L}^{ij} N_i N_j^*$

We get again in the frame where $\bm{p}$ is in the $z$ direction
\begin{equation}
\frac{1}{m_{\tau} m_{\nu}} \left\{ 2 {\bm{p}}^2  \delta_{i3} \delta_{j3} + \delta_{ij} (p \cdot p') \right\} N_i N_j^* =
\frac{1}{m_{\tau} m_{\nu}} \left\{ 2 {\bm{p}}^2  N_0 N_0^* + \left( E_{\tau} E_{\nu} - {\bm{p}}^2   \right) \sum_{\mu} N_{\mu} N^*_{\mu} \right\}.
\end{equation}

\begin{equation*}
N_0 N^*_0 \rightarrow \frac{1}{6} q^2 \frac{1}{4 \pi} \int d \Omega Y_{1,-(M+M')} ({\hat{\bm{q}}}) Y^*_{1,-(M+M')} ({\hat{\bm{q}}}) \Big\{ (-1)^{-M'} \delta_{M0} + 2 (-1)^{-M}
\end{equation*}
\begin{equation*}
 \times \mathcal{C} \left(1 1 1; M,0,M \right) \mathcal{C} \left(1 1 1; M',-M-M',-M \right) \Big\} \Big\{ (-1)^{-M'} \delta_{M0} + 2 (-1)^{-M}
\end{equation*}
\begin{equation}
\label{eq:14-1}
 \times \mathcal{C} \left(1 1 1; M,0,M \right) \mathcal{C} \left(1 1 1; M',-M-M',-M \right) \Big\}.
\end{equation}
Now for the $\delta_{M0} \delta_{M0}$ term we have
\begin{equation}
\sum_{M'} \delta_{M0} \delta_{M0} = 3.
\end{equation}
For the crossed term in Eq.~\eqref{eq:14-1}, $\delta_{M0} \ \mathcal{C} (\cdots) \ \mathcal{C} (\cdots) $, we have
\begin{equation}
\delta_{M0}  \mathcal{C} \left(1 1 1; M,0,M \right) =   \mathcal{C} \left(1 1 1; 0,0,0 \right) = 0.
\end{equation}

The last term in Eq.~\eqref{eq:14-1} involves
\begin{equation}
\sum_{M} \mathcal{C} \left(1 1 1; M,0,M \right)^2 \sum_{M'}  \mathcal{C} \left(1 1 1; M',-M-M',-M \right)^2 = 1.
\end{equation}
Hence, altogether the $N_0 N_0^*$ contribution is
\begin{equation}
\overline{\sum} \sum |t|^2 =  \frac{7}{6} \frac{1}{\pi} \tilde{p}_1^2.
\end{equation}

Next we must evaluate $\sum_{\mu} N_{\mu} N^*_{\mu}$
\begin{equation*}
\sum_{\mu}  N_{\mu} N^*_{\mu} = \frac{1}{6} q^2 \frac{1}{4 \pi} \int d\Omega Y_{1,\mu-(M+M')} ({\hat{\bm{q}}}) Y^*_{1,\mu-(M+M')} ({\hat{\bm{q}}}) \sum_{\mu} \Big\{ (-1)^{-M'} \delta_{M \mu} + 2 (-1)^{-M}
\end{equation*}
\begin{equation*}
\times \mathcal{C} \left(1 1 1; M,-\mu,M-\mu \right)  \mathcal{C}\left(1 1 1; M',-M-M'+\mu,-M+\mu \right) \Big\} \Big\{ (-1)^{-M'} \delta_{M \mu}
\end{equation*}
\begin{equation}
\label{eq:15-1}
+ 2 (-1)^{-M}  \mathcal{C} \left(1 1 1; M,-\mu,M-\mu \right)  \mathcal{C}\left(1 1 1; M',-M-M'+\mu,-M+\mu \right) \Big\}.
\end{equation}

The first term involves
\begin{subequations}
\begin{align}
& \sum_{\mu} \delta_{M \mu} \delta_{M \mu} = \delta_{MM},\\
& \sum_{M,M'} \delta_{M M} = 9.
\end{align}
\end{subequations}

The crossed term involves
\begin{equation}
\sum_{M'} (-1)^{M'} \mathcal{C} \left(1 1 1; M',-M',0 \right) = 0,
\end{equation}
and vanishes, and the product of the second terms in Eq.~\eqref{eq:15-1} gives
\begin{equation*}
\sum_{M M'} \frac{1}{6} q^2 \frac{1}{4\pi} \int d \Omega Y_{1,\mu-(M+M')} ({\hat{\bm{q}}}) Y^*_{1,\mu-(M+M')} ({\hat{\bm{q}}}) 4 \sum_{\mu}  \mathcal{C} \left(1 1 1; M, -\mu, M-\mu\right)^2
\end{equation*}
\begin{equation}
\times \mathcal{C} \left(1 1 1; M', -M-M'+\mu, -M+\mu \right)^2.
\end{equation}
Fixing $M-\mu$ the sum over $M$, and $M'$ of the CGC coefficients gives $1$
, and then
\begin{equation}
\sum_{M-\mu} 1 =3,
\end{equation}
and we get altogether for $\bar{L}^{ij} N_i N_j^*$,
\begin{equation}
\overline{\sum} \sum |t|^2 \rightarrow \frac{1}{m_{\tau} m_{\nu}} \frac{7}{2} \ \tilde{p}_1^2  \frac{1}{\pi}.
\end{equation}
Summing the $N_0N_0^*$ and $N_{\mu} N^*_{\mu}$ terms we find
\begin{equation}
\overline{\sum} \sum |t|^2 = \frac{1}{m_{\tau} m_{\nu}} \frac{1}{\pi} \ \tilde{p}_1^2 \left\{ 2 {\bm{p}}^2  \frac{7}{6} + (E_{\tau} E_{\nu} - {\bm{p}}^2 ) \frac{7}{2} \right\} = \frac{1}{m_{\tau} m_{\nu}} \frac{1}{\pi} \ \tilde{p}_1^2  \frac{7}{2} \left\{ E_{\tau} E_{\nu} - \frac{1}{3} {\bm{p}}^2 \right\},
\end{equation}
and finally, considering the weights for the $M_0$ and $N_i$ parts we get
\begin{equation}
\overline{\sum} \sum |t|^2 = \frac{1}{m_{\tau} m_{\nu}} \frac{1}{\pi} \ \tilde{p}_1^2  \left\{h_i^2 \Big[E_{\tau} E_{\nu} +{\bm{p}}^2  \Big] + \frac{7}{2} \bar{h}^2_i \Big[ E_{\tau} E_{\nu} -\frac{1}{3} {\bm{p}}^2 \Big] \right\} \, .
\end{equation}
\end{itemize}
\end{itemize}


\begin{thebibliography}{99}

\bibitem{isgur89}
Nathan Isgur, Colin Morningstar, and Cathy Reader,  Phys.  Rev. D  {\bf 39}, 1357  (1989).


\bibitem{npb84}
H. K\'uhn, F. Wagner, Nucl. Phys. B  {\bf 236}, 16  (1984).


\bibitem{zpc90}
J. H. K\'uhn, A. Santamaria, Z. Phys. C  {\bf 48}, 445  (1990).


\bibitem{PRe88}
B. C. Barish,  R. Stroynowski, Phys.  Rept.   {\bf 157}, 1  (1988).

\bibitem{Rep2006}
Michel Davier, Andreas H\'ocker, and Zhiqing Zhang, Rev. Mod. Phys. {\bf 78}, 1043 (2006).

\bibitem{PRe2010}
M. Antonelli, D. M. Asner, D. Bauer et al, Phys.  Rept.   {\bf 494}, 197  (2010).

\bibitem{pdg}
 C. Patrignani et al. (Particle Data Group). Chin. Phys. C, {\bf 40}, 100001 (2016).

\bibitem{Ryu}
S. Ryu et al. (Belle Collaboration). Phys.  Rev. D  {\bf 89}, 072009  (2014).

\bibitem{Epifanov}
 D. Epifanov et al. (Belle Collaboration). Phys. Lett. B {\bf 654}, 65 (2007).


\bibitem{Buskulic1}
D. Buskulic  et al. (ALEPH Collaboration). Z. \ Phys.\  C {\bf 74}, 263 (1997).

\bibitem{Buskulic2}
 (ALEPH Collaboration). Z. \ Phys.\  C {\bf 70}, 579 (1996).


\bibitem{Barate}
R. Barate et al. (ALEPH Collaboration).   Eur.\ Phys.\ J.\ C {\bf 1}, 65 (1998).

\bibitem{Inami}
K. Inami et al. (Belle Collaboration). Phys. Lett. B {\bf 672}, 209 (2009).

\bibitem{Arms}
K. Arms et al. (CLEO Collaboration), Phys. Rev. Lett. {\bf 94}, 241802  (2005).

\bibitem{Barate2}
R. Barate et al. (ALEPH Collaboration).   Eur.\ Phys.\ J.\ C {\bf 10}, 1 (1999).

\bibitem{Aubert}
B. Aubert et al. (The BABAR Collaboration), Phys. Rev. Lett. {\bf 100}, 011801 (2008).

\bibitem{delAmo}
P. del Amo Sanchez et al. (BABAR Collaboration), Phys. Rev. D {\bf 83},  032002 (2011).

\bibitem{micu}
L. Micu, Nucl. Phys. B {\bf 10}, 521 (1969).

\bibitem{oliver}
A. Le Yaouanc, L. Oliver, O. Pène, and J. -C. Raynal,  Phys. Rev. D {\bf  8}, 2223 (1973).

\bibitem{close} F. E. Close, An Introduction to Quark and Partons, Academic Press, 1979.

\bibitem{Weinberg}
Steven Weinberg, Phys. Rev. {\bf 112}, 1375 (1958).

\bibitem{Leroy}
C. Leroy, J. Pestieau, Phys. Lett. B {\bf 72}, 398 (1978).



\bibitem{Escribano}
R. Escribano, S. Gonz\' alez-Sol\' is, and P. Roig,  Phys. Rev. D  {\bf 94}, 034008  (2016).

\bibitem{roig}
E.A. Garc\' es, M. H. Villanueva, G. L.  Castro, P. Roig,  J. High Energ. Phys.  {\bf 1712}, 027 (2017).

\bibitem{zou}
A. Mart\' inez Torres, L. S. Geng, L. R. Dai, B. X. Sun, E. Oset and B.S. Zou,
Phys. Lett. B {\bf 680}, 310 (2009).


\bibitem{liang}
W.H. Liang, E. Oset, Phys. Lett. B {\bf 680}, 310 (2009).

\bibitem{bramon}
A. Bramon, A. Grau, G. Pancheri,   Phys.\ Lett.\ B {\bf 283} (1992) 416

\bibitem{itzy}
C.~Itzykson and J.~B.~Zuber,  Quantum Field Theory, Mecraw-Hill, 1980.

\bibitem{bohr}
A.Bohr and B. R. Mottelson, Nuclear Structure, Volume 1. World Scientific, 1998.

\bibitem{mandl}
F. Mandl and G. Shaw, Quantum Field Theory, John Wiley $\&$ Sons, 1984

\bibitem{Gasser}
J. Gasser, H. Leutwyler,  Ann.  Phys.,  {\bf 158}, 142 (1984).

\bibitem{Scherer}
S.~Scherer,  Adv.\ Nucl.\ Phys.\  {\bf 27}, 277  (2003).

\bibitem{rose}
M. E. Rose, Elementary Theory of Angular Momentum, John Wiley $\&$ Sons, 1957.

\bibitem{Sekihara}
Fernando S. Navarra, Marina Nielsen, Eulogio Oset, and Takayasu Sekihara,
Phys. Rev. D {\bf 92}, 014031 (2015).


\bibitem{Ikeno}
N. Ikeno and E. Oset, Phys. Rev. {\bf D 93}, 014021 (2016).

\bibitem{Pelaez:2015qba}
  J.~R.~Pel\'aez,
  Phys.\ Rept.\  {\bf 658}, 1 (2016)
  doi:10.1016/j.physrep.2016.09.001
  [arXiv:1510.00653 [hep-ph]].

\bibitem{Asner}
D. M. Asner et al. (CLEO Collaboration), Phys. Rev. D {\bf 62}, 072006 (2000).

\bibitem{lutz}  M. F. M. Lutz, E. E. Kolomeitsev,  Nucl. Phys. A {\bf 730}, 392 (2004).

\bibitem{rocaSingh}
L. Roca, E. Oset, and J. Singh,  Phys. Rev. D {\bf 72}, 014002 (2005).

\bibitem{geng}
L. S. Geng, E. Oset, L. Roca, and J. A. Oller,  Phys. Rev. D {\bf 75}, 014017 (2007).

\bibitem{molina}
R. Molina, D. Nicmorus, and E. Oset,   Phys. Rev. D {\bf 78}, 114018 (2008).

\bibitem{gengvector}
L. S. Geng and E. Oset, Phys. Rev. D {\bf 79}, 074009  (2009).

\bibitem{a1exp}
S. Schael, The ALEPH Collaboration, R. Barate et al, Phys. Rep. {\bf 421}, 191 (2005).


\bibitem{Leupold}
M. Wagner and S. Leupold,  Phys. Rev. D {\bf 78}, 053001   (2008).

\bibitem{Pich}
D. G\'omez Dumm, A. Pich, and J. Portol\'es,  Phys. Rev. D {\bf 69}, 073002  (2004).


\end{thebibliography}
\end{document}